\documentclass[twocolumn,twocolappendix]{aastex701}
\usepackage{xcolor}
\usepackage{soul}
\usepackage{amsmath}
\usepackage{afterpage}
\usepackage{tabularx}
\usepackage{ltablex}
\usepackage{ulem} 

\let\oldhat\hat
\renewcommand{\vec}[1]{\boldsymbol{#1}} 
\renewcommand{\hat}[1]{\oldhat{\boldsymbol{#1}}}

\newcommand{\simname}[1]{\texttt{#1}}
\newcommand{\orcid}[1]{\href{https://orcid.org/#1}{\includegraphics[width=10pt]{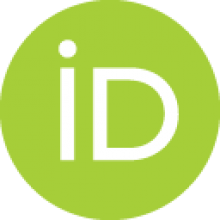}}}
\newcommand{\CB}{r_{\rm CB}}

\newcommand{\CR}{r_{\rm CR}}

\newcommand{\rOTZ}{r_{\rm OTZ}}

\newcommand{\tauG}{\tau_{\rm grav}}
\newcommand{\tauV}{\tau_{\rm visc}}
\newcommand{\CtauG}{\mathcal{T}_{\rm grav}}
\newcommand{\CtauV}{\mathcal{T}_{\rm visc}}

\newcommand{\Mstar}{M_{\ast}}

\newcommand{\EDTZ}{\textup{E\textsc{n}DT\textsc{ran}Z}}

\newcommand{\beq}{\begin{equation}}
\newcommand{\eeq}{\end{equation}}

\newcommand{\Msun}{M_{\odot}}

\received{August 30}
\revised{February 10}
\accepted{February 12}

\shorttitle{Envelope-disk transition zone}
\shortauthors{Das et al. (2025)}

\newcommand{\modelone}{\textsc{MODEL}\textnormal{-1}}
\newcommand{\modeltwo}{\textsc{MODEL}\textnormal{-2}}
\newcommand{\modelthree}{\textsc{MODEL}\textnormal{-3}}

\begin{document}

\title{Modelling the Break in the Specific Angular Momentum within the Envelope-Disk Transition Zone}

\correspondingauthor{Indrani Das, Shantanu Basu, Nagayoshi Ohashi}

\author[orcid=0000-0002-7424-4193]{Indrani Das}
\affiliation{Academia Sinica Institute of Astronomy and Astrophysics, No. 1, Sec. 4, Roosevelt Road, Taipei 106319, Taiwan}
\email[show]{idas2@uwo.ca}

\author[orcid=0000-0003-0855-350X]{Shantanu Basu}
\affiliation{Canadian Institute for Theoretical Astrophysics, University of Toronto, 60 St. George St., Toronto, ON M5S 3H8, Canada}
\affiliation{The University of Western Ontario, Department of Physics and Astronomy, London, ON N6A 3K7, Canada} 
\email[show]{basu@uwo.ca}

\author[orcid=0000-0003-0998-5064]{Nagayoshi Ohashi}
\affiliation{Academia Sinica Institute of Astronomy and Astrophysics, No. 1, Sec. 4, Roosevelt Road, Taipei 106319, Taiwan}
\email[show]{ohashi@asiaa.sinica.edu.tw}

\author[orcid=0000-0002-6045-0359]{Eduard Vorobyov}
\affiliation{Institut für Astro- und Teilchenphysik, Universität Innsbruck, Technikerstraße 25, 6020 Innsbruck, Austria}
\affiliation{Department of Astrophysics, University of Vienna, Tuerkenschanzstrasse
17, 1180, Vienna, Austria}
\email{eduard.vorobiev@univie.ac.at}
\affiliation{Research Institute of Physics, Southern Federal University,
Rostov-on-Don 344090, Russia}

\author[orcid=0000-0002-8238-7709]{Yusuke Aso}
\affiliation{Korea Astronomy and Space Science Institute, 776 Daedeok-daero, Yuseong-gu, Daejeon 34055, Republic of Korea}
\affiliation{Division of Astronomy and Space Science, University of Science and Technology, 217 Gajeong-ro, Yuseong-gu, Daejeon 34113, Republic of Korea}
\email{yaso@kasi.re.kr}


\begin{abstract}
The observations of protostellar systems show a transition in the radial profile of specific angular momentum (and rotational velocity), as evolving from $j\sim{\rm constant}$ ($v_{\phi}\sim r^{-1}$) in the infalling-rotating envelope to $j\propto r^{1/2}$ ($v_{\phi}\sim r^{-1/2}$)  in the Keplerian disk. 
We employ global MHD disk simulations of gravitational collapse starting from a  supercritical prestellar core, that forms a disk and envelope structure in a self-consistent manner, in order to determine the physics of the Envelope-Disk Transition Zone (\EDTZ). 
Our numerical results show that the transition from the infalling-rotating envelope to Keplerian disk happens through a jump in the $j-r$ profile over a finite radial range, which is characterized by the positive local gravitational torques.
The outer edge of the \EDTZ\, is identified where 
the radial infall speed ($v_r$) begins a sharp decline in magnitude and $j$ begins a transition from $j\sim{\rm constant}$ toward $j\sim r^{1/2}$. 
Moving radially inward, the centrifugal radius ($\CR$) is defined where $v_{\phi}$ first transitions to Keplerian velocity at the disk’s edge. Farther inward of $\CR$, model disk develops a super-Keplerian rotation due to self-gravity. The inner edge of the \EDTZ\, is defined at the centrifugal barrier ($\CB$) where $v_r$ drops to negligible values. Inside $\CB$, 
a net negative gravitational torque drives mass accretion onto the protostar. 
On observational grounds, we identify a jump in the observed $j-r$ profile  in L1527 IRS for the first time using the ALMA eDisk data. Comparison with the numerical radial behavior from our MHD disk simulations suggests the observed $j-r$ jump can be used as a kinematical tracer for the existence of $\EDTZ$.
Our results offer insights into the observable imprint of angular momentum redistribution mechanisms during star-disk formation. 
\end{abstract}



\section{Introduction}\label{sec:intro}
Stars form within the dense regions of molecular clouds and the formation of a protostellar disk–star system begins with the gravitational collapse of a dense molecular cloud core.
The protostellar disk plays a fundamental role in regulating mass accretion onto the protostar and in redistributing angular momentum. 
The formation and evolution of the disks are considered to be shaped by gravitational torques \citep[e.g.,][and their series of papers]{VorobyovBasu2007, VorobyovBasu2008,VorobyovBasu2009}, nonideal MHD effects, rotation–magnetic field misalignment, and turbulence \citep[for reviews, see e.g.,][]{Li+2014PP6,Tsukamoto+2023PPP7}. 
A central question for such deeply embedded protostellar disks is their connection to the infalling envelope, which may govern the 
dynamical evolution 
at their envelope–disk transition.
Recent observational evidences, such as L1527 IRS shows a break in the radial profile of rotational velocity ($v_{\phi}-r$) 
at the envelope-disk transition  as going from $r^{-1}$ to $r^{-1/2}$ \citep{Ohashi+2014,Aso+2015, Aso+2017,Hoff+2023}, that has emphasized the 
importance of angular momentum redistribution during
the formation of a protostellar disk-star system. 
This characteristic break in $v_{\phi}-r$ at the envelope-disk transition is also associated with the observed break in the radial profile of specific angular momentum ($j-r$) 
at the interface of the infalling-rotating envelope and the outward moving edge of the Keplerian disk.

It is crucial to understand whether there exists a physical transition zone at the envelope-disk interface where the infall velocity within the envelope may gradually decelerate over an extended region
rather than abruptly impacting the rotationally supported disk such that  the transition in $v_{\phi}$ from $\propto r^{-1}$ to $\propto r^{-1/2}$ gradually happens over a finite radial zone.
Numerical studies of infall dynamics at the envelope–disk transition \citep[e.g.,][]{Jones+2022,Shariff+2022}, based on hydrodynamic simulations, demonstrate that the physics at this interface is far more complex than predicted by a classical test-particle approach (see Methods Section of \citealt{Sakai+2014}, Eq. (2)-(4) of \citealt{Shariff+2022}).
According to the classic picture of the motion of a (non-interacting)
test particle in the gravity field of a central stellar object of a fixed
mass under the constraint of energy and angular momentum conservation, 
the centrifugal barrier ($\CB$) is defined where the kinetic energy of the infalling gas is converted to rotational energy, bringing the infall to a complete halt as derived from the conservation of angular momentum and energy under a central force potential.
It is considered to be located at a radius equal to half of the centrifugal radius $\CR$, $\CB = \CR /2$ (see  
    Fig.~1 of \citealt{Terebey+2025}, \citealt{Sakai+2014}).
The centrifugal or Keplerian radius ($\CR$) is defined where the rotational speed of the test particle
reaches the (local) Keplerian speed as the centrifugal force exactly balances the local gravitational force at that radius.
However, under hydrodynamic (HD) or magnetohydrodynamic (MHD) fluid treatments, this may be subject to change as a result of the interplay among various physical mechanisms, such as gas pressure, self-gravity, magnetic fields, etc. 
In our numerical study,  
we use the term ``centrifugal barrier" to describe the region where the radial infall speed drops to negligible values; however, this is not the same as the centrifugal barrier defined from energy and momentum conservation from the classical test particle approach.

In this study, we carry out global nonaxisymmetric magnetohydrodynamic disk simulations starting
from a rotating supercritical prestellar core collapse to investigate the observed break in the radial profile of specific angular momentum at the envelope-disk transition and determine the physics of such a radially extended envelope-disk transition zone (\EDTZ). 
Starting from such an initial condition results in the self-consistent formation of a protostellar disk-star system. 
Thereafter, to strengthen the observational evidence of \EDTZ, we re-investigate L1527 IRS system, which showed the transition from the infalling-rotating motion to Keplerian motion through an observed $j-r$ (and $v_{\phi}-r$) jump that has been first identified in our study, based on the C$^{18}$O data obtained in the ALMA Large Program eDisk  observations 
\citep[Embedded disks in early planet formation;][]{Ohashi+2023,Hoff+2023}. 
We construct the observed $j-r$ diagram from the observed rotation curve to visually identify  a gradual transition (appeared as a jump) from the
infalling envelope  to outward moving edge of the Keplerian disk of L1527 IRS. 
In the first eDisk study of L1527 IRS by \cite{Hoff+2023}, the rotation curve of the L1527 IRS has been  presented. However, the possible existence of the \EDTZ\, 
and its connection to the observed $j-r$ jump has been  first recognized in our study. 
We aim to
build a conceptual framework of \EDTZ\, by examining its self-consistent development, complemented by the observational signatures of a $j-r$ jump.

The paper is organized as follows. 
In Sec. \ref{sec:methods}, we describe the numerical setup. 
We present our results in Sec. \ref{sec:results}. 
We illustrate the modeling of disk evolution, with an emphasis on the physics of the \EDTZ\, and the development of the jump in the $j-r$ and $v_{\phi}-r$ profile in Sec. \ref{sec:theory}. 
Then in Sec. \ref{sec:obs}, we present the $v_{\phi}-r$ and $j-r$ diagram of L1527 IRS 
based on the re-investigations of the original eDisk data, focusing on the identified $j-r$ jump at the envelope-disk transition and its qualitative comparison with the radial behavior found in our numerical results.
In Sec. \ref{sec:discussion}, we discuss our results in relation to
contemporary studies, elucidating the significance of the $j-r$ jump as an indicator of the \EDTZ\, and summarize our key findings in Sec. \ref{sec:conslusions}. 
Throughout this paper, we extensively use the $j-r$ jump in explaining our results because the $j-r$ jump appears to be visually more pronounced than the $v_{\phi}-r$ jump.

\section{Methodology}   \label{sec:methods}
We perform two-dimensional numerical global magnetohydrodynamic disk simulations with self-consistent cloud-to-disk transition starting from the gravitational collapse of a supercritical prestellar cloud core, under the conditions of a weak magnetic field in the ideal MHD limit using the FEOSAD (Formation and Evolution Of a Star And its circumstellar Disk) code \citep[][]{Vorobyov+2020a}. 
The \textsc{FEOSAD} code follows the thin-disk limit.
We study the formation and evolution of protoplanetary disks (PPDs) in low-mass stars for an initial prestellar core with a mass of $1\,\Msun$. 
In our numerical model, the prestellar cloud core has the form of a flattened magnetized pseudodisk, a spatial configuration
that can be expected in the presence of rotation and large-scale magnetic fields \citep{Basu1997}.  
The effects of magnetic fields are modeled using the flux-freezing approximation with the constant dimensionless mass-to-magnetic flux ratio ($\mu$) of 10. 
The main physical processes considered in \textsc{FEOSAD} code
include the self-gravity of disk,
magnetic field pressure and tension, 
viscous transport, 
including the key thermal processes such as stellar irradiation, viscous heating, compressional heating (i.e., $PdV$ work), dust thermal cooling 
(or the radiative cooling 
from the disk surface), all of which are needed to provide a realistic temperature structure of the disk.
The governing equations of the existing physical processes adopted in \textsc{FEOSAD} code, including the magnetic field geometry appropriate for the thin-disk approximation 
in the code and the definition of $\mu$ are presented in Appendix \ref{sec:gas}. 
\textsc{FEOSAD} also models the occurrence of mass accretion bursts
in a gravitationally (GI) unstable magnetized PPD with globally suppressed but episodically
triggered magnetorotational instability (MRI). 
Although not in the focus of the current work, this module is an integral part of FEOSAD and we refer to Appendix \ref{sec:app3} for a discussion of the numerical treatment of MRI activation.

The simulation is initialized by the initial profile of gas surface density and an initial angular velocity profile consistent with an axially symmetric core collapse model \citep[]{Basu1997}, refer to Appendix \ref{sec:app1}. 
We note that the adopted form of the column density is very similar to the integrated column density of Bonnor–Ebert sphere \citep{DappBasu2009,DasBasu2022}.
For the boundary conditions, refer to Appendix \ref{sec:app2}.  
After a central protostar forms (at a moment labeled $t=0$), a centrifugally supported disk is formed when the  inner infalling and spinning-up layers of the core hit the centrifugal barrier near the central sink cell. 
The sink cell is introduced at ${r_{\rm sc}}$ =0.52~au to avoid too small time steps
imposed by the Courant condition. 
In the continuing evolution, the central star is surrounded by a rotationally supported accretion disk along with the remaining infalling envelope. 
Our global disk formation simulations employ one of the smallest possible sink cells, roughly consistent with the thin-disk simulations of \citet{Bae2014}, where an inner sink radius of 0.2~au is used.

Our current numerical setup of \textsc{FEOSAD} code employs polar coordinates on a two-dimensional  grid with $256 \times 256$ grid zones. 
The radial grid is logarithmically spaced, yielding a minimum radial grid size ($\Delta r$) of $\sim\,1.91\times 10^{-2}$ au near the inner boundary and increasing up to $\sim$3.62 au at the disk's edge ($\sim$ 100~au), which covers the extent required for the disk evolution by the desired timeline. 
The azimuthal
grid is equispaced, that yielding a minimum polar grid size ($r \Delta \phi$) of $1.29\times 10^{-2}$ au near the inner boundary and increasing up to 2.52 au at the disk's edge.
In our numerical model, the \EDTZ\, is resolved in the radial direction by approximately 18 and 11 grid zones at two key evolutionary times, $t=$ 44.2 kyr and 73 kyr, respectively, at which an extended analysis of it is presented in Sec. \ref{sec:theory}.

In \textsc{FEOSAD}, the disk gravitational potential is computed using the integral form of the gravitational potential using the convolution theorem \citep[see Appendix B in][]{Vorobyov+2024}. 
In our numerical setup, gravitational torques play a principal role in
inward mass transport during the embedded phase at a rate that is in agreement with typical accretion rates inferred in \cite{Hartmann1998}. 
In FEOSAD code, viscous torques also contribute to the mass transport within the disk. 
For the detailed description and self-consistent calculation of gravitational and viscous torques, we refer to \cite{VorobyovBasu2007} and \cite{VorobyovBasu2009}, respectively.
Furthermore, the torques due to artificial viscosity, as present in our
Eulerian numerical code, are much smaller than the gravitational torques \citep[see][]{VorobyovBasu2007}, meaning that they have little influence on the radial transport of mass in our numerical
simulations. 
In addition, the use of the third-order accurate flux interpolation scheme \cite{ColellaWoodward1984} guaranties low numerical diffusion.

In the current study,
we perform three numerical models \modelone, \modeltwo, and \modelthree, each having identical initial conditions but distinct values of
the initial background turbulent viscosity of  $10^{-4}$,
$10^{-3}$, $10^{-2}$, respectively. 
We refer to Table \ref{tab:table} for the initial values of all the model parameters. 
The initial background adaptive viscosity across the disk of the order of $\alpha_{\rm visc}=10^{-4}-10^{-3}$ (refer to Eq. \ref{eq:alphaeff} in Appendix \ref{sec:app3}), is in accordance with recent observations of efficient dust settling in protoplanetary disks \citep{Rosotti2023}. 
In our study, \modeltwo~ serves as the fiducial model for facilitating the comparison between lower-viscosity and higher-viscosity model.

In our numerical treatment, the protostellar disk occupies the inner
part of the numerical polar grid and its outer  parts are exposed to intense mass-loading from the infalling core in the initial embedded phase of evolution.
We note that in the thin-disk approximation all material from
the envelope lands onto the outer disk regions. 
This is a reasonable assumption according to the demonstration presented by \cite{visser+2009}, in which the gas trajectories in  the infalling envelope 
are accurately calculated and it is found that the bulk of the infalling material was deposited onto the outer edge of the disk. 
A recent study by \cite{RedkinVorobyov2025} 
also showed that about 50\% of the envelope mass falls onto the very narrow annulus of the disk near disk’s inner edge in the framework of \cite{CassenMoosman1981} model. 
This is valid under the approximation when the flaring of the disk is not taken into account \citep[refer to Fig. 11 and \S Appendix A of][]{VorobyovBasu2010}. 
It is also noted that our disk model is not razor-thin and we account for the disk vertical extent by computing the vertical scale height of gas in each computational cell, which is further used in the relevant calculations, for example, for the heating function when accounting for the stellar radiation flux intercepted by the disk surface, etc.  \citep[see further in][]{Vorobyov+2020b}.


\section{Results}   
\label{sec:results}
In this section, we present the dynamical evolution of several physical characteristics during 
the star-disk formation, with a focus on the development of the transition zone at the envelope-disk interface as obtained from our numerical simulations (refer to Sec. \ref{sec:theory}). 
We delve into the results obtained from \modeltwo~ (fiducial model), with brief comparisons to the lower- and higher- viscosity counterparts \modelone ~and \modelthree~, respectively (see Table \ref{tab:table}). 
In Sec. \ref{sec:obs}, we present  eDisk observations of L1527 IRS and compare them with our numerical models, with an emphasis on their radial behavior within the \EDTZ.

\begin{table*}[!ht]
\centering
\begin{tabular}{ccccccccccccc}
    \hline \\
    Model & $M_{\rm core}$ & $r_{\rm out}$ & $r_{\rm in}$ & $r_0$ & $\Sigma_{g,0}$ & $\Omega_0$  &  $\beta$ [\%] & $\lambda$ & 
   $\alpha_{\rm MRI}^{\rm}$ & $\alpha_{\rm MRI}^{\rm Burst}$ 
    & $\alpha_{\rm dz}^{\rm}$ 
    & $M_{*,{\rm f}}$ \\ 
    & [$M_{\odot}$] &  [au] &  [au] &  [au] & [g.${\rm cm}^{-2}$]  & $[{\rm km} \ {\rm /s /pc}]$ & & & & & &[$M_{\odot}$] \\ 
    \hline \\
     \simname{MODEL-1} & 1.01 & $5053.5$ & 0.52 & $630.05$ & $5.08\times10^{-1}$ & 2.00  & $1.33\times10^{-3}$ & 10.0 & $ 10^{-4}$ & 0.1 & $10^{-5}$ & 0.57\\
     \simname{MODEL-2} & 1.01 & $5053.5$ & 0.52 & $630.05$ & $5.08\times10^{-1}$ & 2.00  & $1.33\times10^{-3}$ & 10.0 & $ 10^{-3}$ & 0.1 & $10^{-5}$ & 0.58\\
     \simname{MODEL-3} & 1.01 & $5053.5$ & 0.52 & $630.05$ & $5.08\times10^{-1}$ & 2.00  & $1.33\times10^{-3}$ & 10.0 & $ 10^{-2}$ & 0.1 & $10^{-5}$ & 0.62\\
     \hline \\
\end{tabular}
\vspace*{-0.3cm}
\caption{Model parameters (from left to right): $M_{\rm core}$ is the initial core mass, $r_{\rm out}$ is the initial outer radius of the prestellar core, ${\rm r_{in}}$ is inner boundary of the computational domain, $r_0$ is the radius of the central plateau in the initial core, $\Sigma_{\rm g,0}$ is the initial gas surface density at the center of the core , 
$\Omega_0$ is initial angular speed of the prestellar rotating core, $\beta$ is initial ratio of the rotational to gravitational energy, $\lambda$ is the initial mass to magnetic-flux ratio normalized to a factor of $(2\pi\sqrt{G})^{-1}$,   
$\alpha_{\rm MRI}$ is the constant background turbulent viscosity across the disk extent, $\alpha_{\rm MRI}^{\rm Burst}$ is the maximum turbulent viscosity at the occurrence of an MRI outburst only, $\alpha_{\rm dz}$ is the nonzero residual viscosity parameter in the MRI-dead layers in the absence of the burst,   
$M_{*,{\rm f}}$ is the final stellar mass at a desired timeline of $\sim100.0 \ {\rm kyr}$.} 
\label{tab:table}
\end{table*}

\subsection{Theoretical aspects of the transition zone} \label{sec:theory}

\begin{figure}[!htb]
\includegraphics[width=\linewidth]{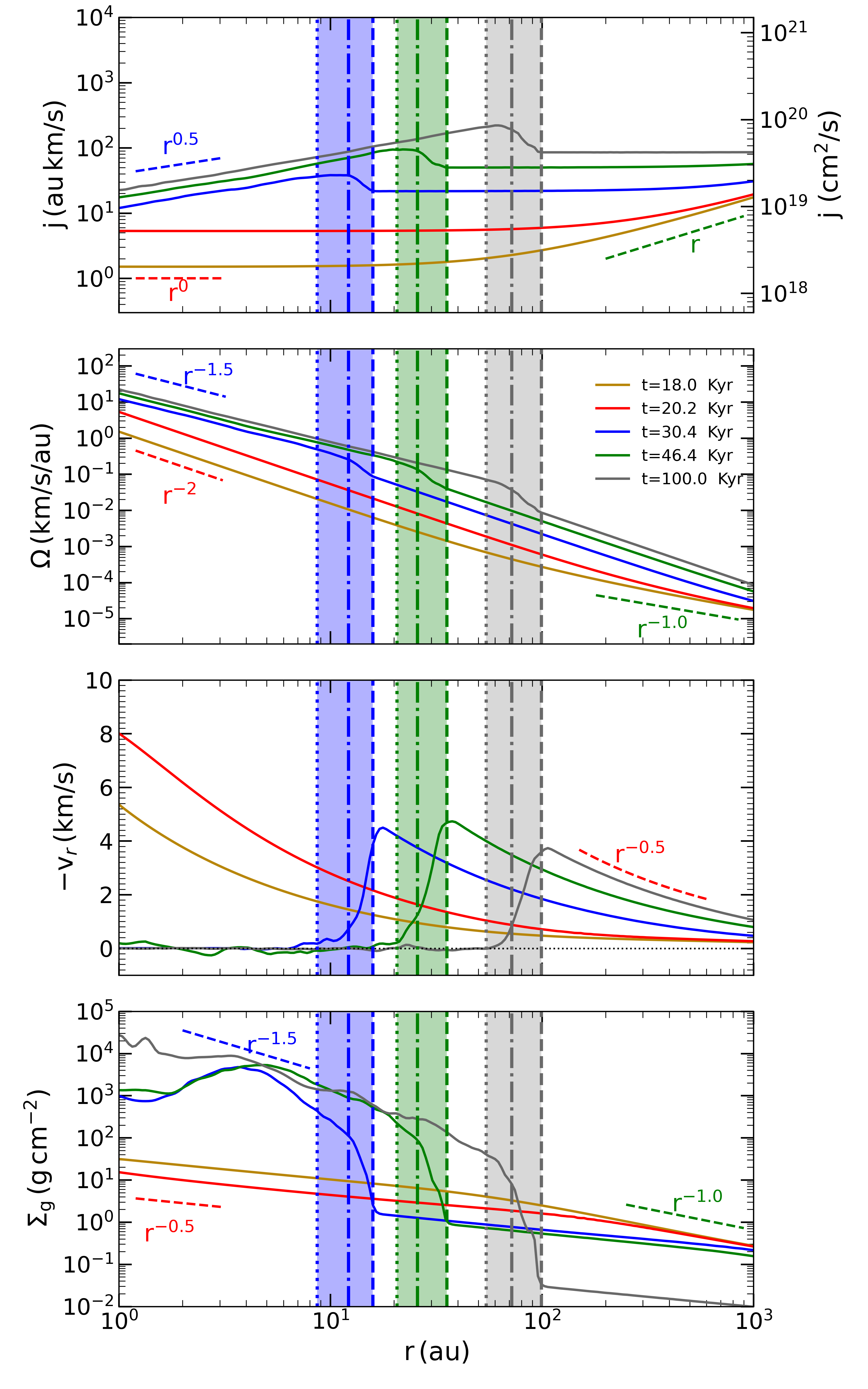}
\vspace*{-0.8cm}
\caption{Radial profiles of the azimuthally averaged quantities in the equatorial plane obtained from \modeltwo~ (from top to bottom): specific angular momentum ($j$), angular velocity ($\Omega$), infall velocity ($v_{r}$), and gas surface density ($\Sigma_{\rm g}$) at different evolutionary times. The evolutionary tracks exhibiting 
$j \sim r^{1/2}$ represent the evolutionary stages after the disk formation. 
The respective colored vertical dashed, dash-dotted, and dotted line respectively represent $\rOTZ$, $\CR$, and $\CB$, respectively, corresponding to the respective evolutionary times.
The colored vertical strips present the \EDTZ s at the respective timestamps. 
Each \EDTZ\, is bounded by the $\rOTZ$ and  $\CB$. 
}
\label{fig:figradial}
\end{figure}

\begin{figure*}[!htb]
\centering
\vspace*{-0.2cm}
\includegraphics[width=0.9\linewidth]{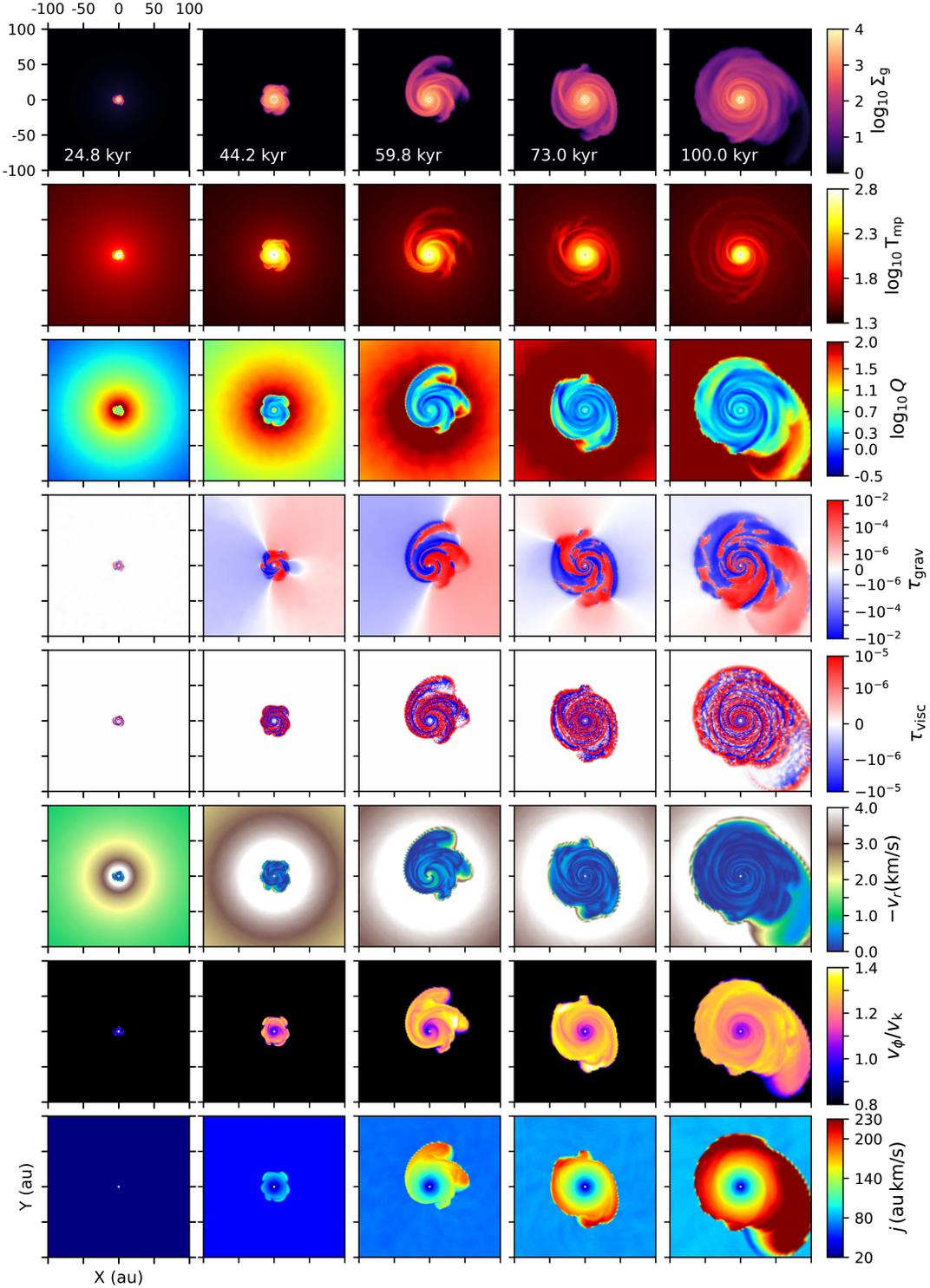}
\vspace*{-1.0cm}
\caption{Evolution of the gas surface density ($\Sigma_{\rm g}$, in units of $\log_{10} \, {\rm g \, {cm}^{-2}}$), midplane temperature ($T_{\rm mp}$, in units of $\log_{10} \, {\rm K}$), Toomre-$Q$ parameter, local gravitational torque ($\tauG$) and local viscous torque ($\tauV$) in code units (the conversion factor is $=7.79\times 10^{40}\, {\rm dyne\,cm}$),  infall speed ($-v_r$ in units of km/s), rotational velocity in  units of Keplerian velocity ($v_{\phi}/v_{\rm K}$), and specific angular momentum ($j$ in  units of au km/s), 
shown  over a region of $100 \times 100 \, {\rm au}^2$ in the midplane of the disk, showing the large-scale disk structure obtained from \modeltwo~ at distinct time instances after the onset of collapse.}
\label{fig:2DEvolDisk}
\end{figure*}

Fig. \ref{fig:figradial} shows the radial profiles of specific angular momentum ($j$), angular velocity ($\Omega$), infall velocity ($v_r$), and gas surface density ($\Sigma_{\rm g}$) at distinct evolutionary times  as obtained from \modeltwo~ (refer to Fig. \ref{fig:2DEvolDisk} for the two-dimensional illustration of the model disk). 
For a self-gravitating isothermal cloud that slightly
exceed the maximum limit allowable for hydrostatic equilibrium, tend to develop a surface density profile of the form $\Sigma_{\rm g} \propto r^{-1}$ in the nearly
static outer envelope as long as the initial conditions allow the
early phases of the flow to evolve at nearly sonic speed. 
Such radial behavior of $\Sigma_{g}$
is analogous to the volume density profile of the form $\rho \sim r^{-2}$ for a spherical isothermal collapse \citep[][]{BodenheimerSweigart1968,Shu1977}.
As the inside out collapse commences, the point-mass-like protostar forms at the center, the surface density profile takes the form $\Sigma_{\rm g} \propto r^{-1/2}$, which corresponds to the conventional self-similar profiles for the freely falling regions inside a radial expansion wave \citep[equivalent to $\rho\sim r^{-3/2}$ for the case of a spherical collapse, refer to][]{Shu1977,Das+2025a}. 
Within the expanding region of the radial
expansion wave, the infall speed approaches
free fall ($v_r \sim r^{-1/2}$). 
Before the formation of the protostellar disk,
the specific angular momentum is  spatially quasi-uniform ($j\sim {\rm constant}$) within the inner infalling-rotating envelope region  and evolves as $j\propto r$ in the outer static region. 
During the protostar-dominated stage, mass is brought in from the envelope, and angular momentum is carried along with it. As the stellar mass grows with time, the levels of the $j-r$ plateau correspondingly rises with time.

Next, after the disk formation at $t$~=22.2 kyr, the specific angular momentum makes a transition from $j\sim {\rm constant}$ to $j\sim r^{1/2}$  through a jump   
that extends across a radial thickness and is embedded within the \EDTZ\, as shown in  Fig. \ref{fig:figradial}.
It is noteworthy that the jump in the $j-r$ profile acts as a critical indicator of such a physical transition zone that lies
between the inner edge of the infalling-rotating envelope and the outward moving edge of the Keplerian disk. 
The \EDTZ\, is indicated by the pink vertical strip spanning between an outer and inner edge of the transition zone. 
We refer to Fig. \ref{fig:torqueanvel_model2} for a detailed discussion on the physics of the \EDTZ. 
In our study, the outer edge of the transition zone ($\rOTZ$) is defined where $j$ begins a transition from $j\sim {\rm constant}$ to $j \sim r^{1/2}$ and the infall speed ($v_r$) starts to decelerate. 
Whereas, the inner edge of the transition zone is defined at $\CB$ where the infall speed ($v_r$) drops to negligible values. 
After the disk formation, the $j\propto r^{1/2}$ levels also rise with time as $\Mstar$ increases with time via disk accretion, causing the Keplerian velocity to increase self-consistently and the disk to rotate faster and the cloud core layers with progressively higher angular momentum approach the disk and accrete on to it as cloud collapse proceeds.

Furthermore, we notice that the radial profile of angular velocity ($\Omega$) exhibits a transition from $\Omega \propto r^{-2}$ (or rotational velocity $v_{\phi} \propto r^{-1}$) in the freely infalling envelope to $\Omega \propto r^{-3/2}$ (or $v_{\phi} \propto r^{-1/2}$) within the Keplerian disk through a jump as shown in Fig \ref{fig:figradial}. 
The jump in $v_{\phi}-r$, and thus in $j-r$ profile, is a manifestation of the inherent jump in $\Omega-r$ at the envelope-disk interface across a radial thickness.
The radial profile of the gas surface density within the disk gradually develops a much steeper profile of the form $\Sigma_{\rm g} \propto r^{-3/2}$, which has the same logarithmic slope as estimates of the minimum mass solar nebula (MMSN) that are made by adding the
solar abundance of light elements to each planet and spreading the
mass through zones surrounding the planetary orbits \citep{Weidenschilling1977}.  
We also note that, a surface density profile of the form of $\Sigma_{\rm g} \sim r^{-3/2}$ is expected for
a gravitationally unstable disk kept at a marginally unstable, self-regulated state with Toomre-$Q\sim 1$ \citep{VorobyovBasu2007}, where $Q=c_s \Omega/(\pi G \Sigma_{\rm g})$ and $c_s$ and $G$ are the local sound speed and gravitational constant, respectively. 
Moreover, the infall velocity ($v_r$) sharply declines in magnitude
over this radially extended \EDTZ. 
Eventually, the radial infall speed drops to negligible values at a radial location, defined as the $\CB$ in our model. 
It is salient that on either side of the \EDTZ, each of the above kinematic quantities exhibits a distinct radial dependence, while across the \EDTZ, they undergo a morphological transformation in a self-consistent manner.

Fig.~\ref{fig:2DEvolDisk} shows the global view of the long-term evolution of a protoplanetary disk since the gravitational collapse of a prestellar core, where the gas surface density, midplane temperature, Toomre-$Q$ parameter are presented in the top three rows over a region of 100 au $\times$ 100 au as obtained from \modeltwo. 
During the embedded phase, by $t\lesssim 100\,{\rm kyr}$, the disk quickly grows up to a radius of $\approx 100 \, {\rm au}$. 
The disk growth is predominantly regulated by mass gain from the collapsing core and mass loss via protostellar accretion. 
We refer to Fig. \ref{fig:1Dmodel} in Appendix \ref{sec:AppSimulations} for the temporal evolution of the mass accretion quantities for \modeltwo.
The spatial map of midplane temperature
of the disk shows that inner regions of the disk are warmer than the outer region.
The spiral structures are clearly imprinted in the temperature distribution, reflecting localized heating along the arms. 
Next, the Toomre-$Q$ parameter compares the destabilizing effects of self-gravity in the disk against the stabilizing effects of pressure and rotation. 
The spatial map of Toomre-$Q$ shows that same asymmetric spiral
structures imprinted in the 2D maps of $\Sigma_{\rm g}$. 
The local disk regions corresponding to Toomre-$Q\lesssim2$ ($\sim0.3$ in its logarithmic units) represents the gravitationally unstable region.

The middle rows of Fig. \ref{fig:2DEvolDisk} present the corresponding spatial maps of local gravitational and viscous torque. 
The local gravitational torque ($\tauG$) can provide physical insight
into the inhomogeneity of local gas surface density ($\Sigma_{\rm g}$). 
In
general, the inhomogeneities that are characterized by negative $\tauG$
are losing angular momentum and spiraling onto the protostar,
while the inhomogeneities that are characterized by positive $\tauG$
are gaining angular momentum and are moving radially outward. 
In the spatial maps of local viscous torque $\tauV$, the 
local disk regions characterized by the negative and positive $\tauV$ indicate inward and outward mass transport within the disk, respectively.
It is noticed that the magnitude of the viscous torques is roughly two orders of magnitude lower than that of the gravitational torques in the embedded phase. 
In our study, we notice that for a global $\alpha_{\rm visc}$ in the range of $10^{-4}-10^{-2}$, the inward mass transport across the disk during the embedded phase is primarily provided by vigorous disk gravitational instability (GI), which is manifested by the appearance of irregular spiral structures.
We refer to the radial torque profiles in Fig.~\ref{fig:torqueangvel_model1model3} in Appendix \ref{sec:AppSimulations} that shows during this early evolution phase the $\tauG$ still dominates $\tauV$ for the other two models.

The last three rows of Fig. \ref{fig:2DEvolDisk} show the spatial maps of radial infall speed ($-v_r$), rotational velocity in units of Keplerian velocity, and specific angular momentum ($j$) at the corresponding evolutionary times. The radial infall speed substantially drops at the disk's outer edge. 
The 2D map of $v_{\phi}$ shows the rotation is sub-Keplerian in the infalling-rotating envelope and the existence of the super-Keplerian region ($v_{\phi}/v_{\rm K} > 1$) within the \EDTZ\, and the disk before $v_{\phi}$ goes back to the Keplerian speed in the innermost region of the disk. 
We refer to Fig. \ref{fig:torqueanvel_model2}d for the detailed discussions on the super-Keplerian region. The \EDTZ\, can essentially be recognized from the 2D maps of $j$ as we notice the jump in $j$ across all polar rays as it transitioning from the envelope to the disk region.

\begin{figure*}[!htb]
\centering
\includegraphics[width=13cm, height=20cm]{combinedv2model2.png}
\vspace*{-0.2cm}
\caption{Azimuthally averaged radial profiles of several physical characteristics at two distinct evolutionary times obtained from \modeltwo~ (from top to bottom): (a) local gravitational ($\tauG$) and viscous ($\tauV$) torque in code units; (b) cumulative (or radially integrated) gravitational ($\CtauG$) and viscous ($\CtauV$) torque in code units; (c) specific angular momentum ($j$); (d) rotational velocity ($v_{\phi}$) and its mass-weighted value ($v_{{\phi},{\rm m-wt}}$); (e) $Q_{\rm min}$ (defined as the minimum in Toomre-$Q$ across all azimuths at each radius rather than an azimuthally averaged quantity) and mass-weighted Toomre-$Q$ parameter $Q_{\rm {m-wt}}$ along with the azimuthal scatter of Toomre-$Q$ at a given radial distance marked by the blue shaded area; (f) infall speed ($v_r$) are presented. 
The horizontal  black dotted line in the top two panels (a \& b) and in the bottom panel (f) present the reference lines for the zero torque and zero infall speed, respectively.
The vertical dashed, dash-dotted, and dotted line represent $\rOTZ$, $\CR$, and $\CB$, respectively. The pink shaded vertical strip represents the \EDTZ\, bounded by $\rOTZ$ and $\CB$. 
The lighter and darker gray shaded regions represent the envelope and the disk interior, respectively.
}
\label{fig:torqueanvel_model2}
\end{figure*}

\begin{figure*}[!htb]
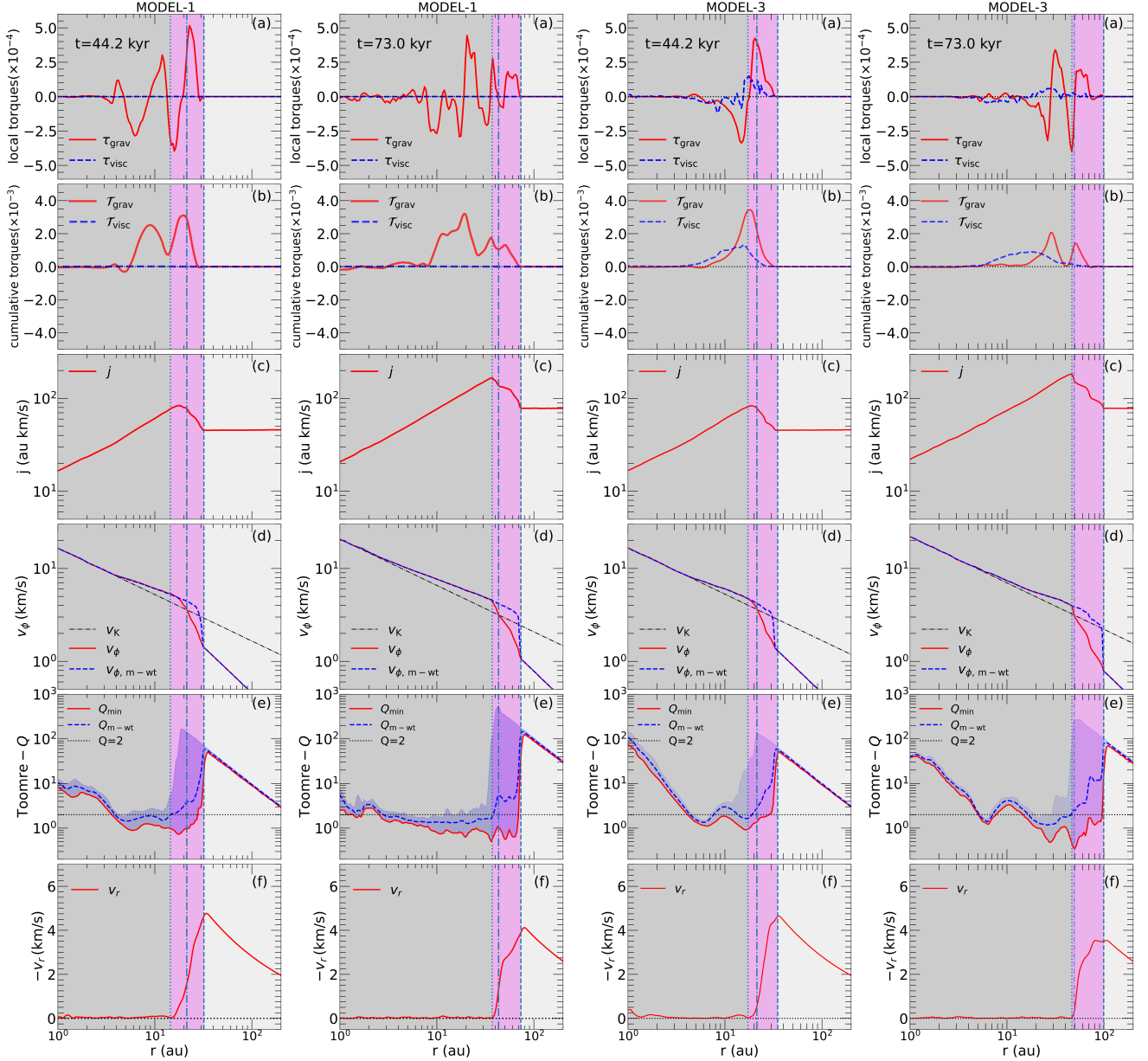

\centering
\includegraphics[width=0.48\linewidth]{combinedv2model1.png}
\includegraphics[width=0.48\linewidth]{combinedv2model3.png}
\caption{Same as in Figure \ref{fig:torqueanvel_model2}, but for \modelone~ (left panel) and \modelthree~ (right panel).}
\label{fig:torqueangvel_model1model3}
\end{figure*}

In Fig.~\ref{fig:torqueanvel_model2}, we explain the physical origin of the \EDTZ\, from our analysis focusing on the internal torques and kinematics at two distinct evolutionary times as obtained from \modeltwo. 
Fig. \ref{fig:torqueanvel_model2}a,b 
show the azimuthally averaged radial profiles of the local gravitational ($\tauG$) and viscous ($\tauV$) torque and the radially integrated (or cumulative) gravitational ($\CtauG$)  and viscous ($\CtauV$) torque (refer to Eq. \ref{eq:tauG}-\ref{eq:tauV}). 
From Fig. \ref{fig:torqueanvel_model2}a, it is noticed that the local gravitational torque ($\tauG$) is virtually zero within the infalling envelope. 
Moving radially inward, the local gravitational torque ($\tauG$) first becomes positive at $\rOTZ$, the outer edge of the \EDTZ, as presented by the vertical dashed line. 
Within the \EDTZ, $\tauG$ remains positive, meaning that the angular momentum is transported inward. 
Thereafter, $\tauG$ declines and it becomes predominantly negative inside the disk, meaning that the angular momentum is transported outward. 
Next, we examine the cumulative gravitational torques ($\CtauG$) exerted as a function of radius. 
Here, $\CtauG$ is calculated by radially integrating the local gravitational torque ($\tauG$) in an outside-in manner, over a bin extending from 100~au to an inner radius. 
We notice that $\CtauG$ increases within the transition zone where $\tauG$ is predominantly positive.  
However, inside the disk, $\CtauG$ decreases as $\tauG$ becomes predominantly negative. 
This radial behavior of the gravitational torques is consistent with the findings of \cite{VorobyovBasu2007,VorobyovBasu2009}.
Note that, the cumulative gravitational torque
$\CtauG$ expresses the efficiency of angular momentum and mass redistribution in the disk as a whole. 
A nonzero net negative $\CtauG$ within the inner disk region indicates the a dominant inward mass transport (or in other words a dominant outward transport of specific angular momentum).
The azimuthally averaged local viscous torque ($\tauV$) remains significantly lower than $\tauG$ within the entire disk region as also seen in Fig.~\ref{fig:2DEvolDisk}. 
Likewise, the cumulative viscous torque ($\CtauV$) is substantially smaller than the $\CtauG$.

The redistribution of the local gravitational torque leads to a corresponding redistribution of angular momentum within the disk. 
As $\tauG$ goes from positive to negative when moving radially inward, it implies that the predominant outward mass transport within the \EDTZ\, transitions into predominant inward mass transport in the disk interior.
The predominant negative local gravitational torque inside the disk redistributes the angular momentum from the inner to the outer region. 
This redistribution of the angular momentum from inner to outer region of the disk and the 
positive gravitational torque within the \EDTZ\, gives rise to the development of the jump in the $j-r$ profile between the spatially uniform specific angular momentum ($j\sim {\rm constant}$) and the Keplerian angular momentum  $j\sim r^{1/2}$, as seen from the azimuthally averaged $j-r$ profiles in Fig. \ref{fig:torqueanvel_model2}c. 
This 
positive $\tauG$ within the \EDTZ\, makes the jump in the $j-r$ profile appear more well defined.  
We refer to Fig \ref{fig:jrpolray} in Appendix \ref{sec:AppSimulations} for the appearance of $j-r$ jump at different polar rays.
From our numerical results, we
notice that during the embedded phase the gravitational torques play the leading role in controlling disk formation and evolution, consistently
surpassing viscous torques. 
For our extended analysis of the \EDTZ\, in two additional models, \modelone\, and \modelthree\, with a global $\alpha_{\rm visc}=$ $10^{-4}\,\, {\rm and}\,\, 10^{-2}$, respectively, we refer to Fig. \ref{fig:torqueangvel_model1model3}.

Our analysis further shows that the jump in the $j-r$ profile at the envelope-disk transition, as seen in Fig. \ref{fig:torqueanvel_model2}c, is attributed to the development of an intermediate power-law profile in $v_{\phi}-r$ profile, 
bridging the $v_{\phi}\sim r^{-1}$ behavior in the infalling–rotating envelope and the $v_{\phi}\sim r^{-1/2}$
behavior in a Keplerian disk, as seen from the azimuthally averaged $v_{\phi}-r$ profiles in Fig. \ref{fig:torqueanvel_model2}d. 
Our model curves of $j-r$ and $v_{\phi}-r$ show that there exists three dynamical
regions: (i) the infalling-rotating envelope as marked by the light gray shaded area, (ii) the \EDTZ\, as marked by pink colored zone, which is an extended radial zone characterized by the $j-r$ jump and the intermediate power-law slope in $v_{\phi}-r$, and (iii) the near-Keplerian disk as marked by the dark gray shaded area.
The corresponding azimuthally averaged radial profile of infall velocity ($v_r$) is presented at the respective timestamps in Fig. \ref{fig:torqueanvel_model2}f.
We notice that infall occurs at the speed of nearly free-fall in the inner envelope, which is consistent with the conditions of rapid collapse of a highly supercritical core. 
It is interesting to note that infall gradually begins to decelerate at $\rOTZ$.  
Thereafter $v_r$ drops to negligible values
at $\CB$.

In Fig.~\ref{fig:torqueanvel_model2}d, it is noticed that in the model disk, we identify the centrifugal radius ($\CR$), where the rotational velocity
first transitions to the Keplerian velocity ($v_{\rm K}$) at the disk's edge. 
Inward
of this centrifugal radius, the rotation transitions to a super-Keplerian value accounting for the disk’s self-gravity, before returning to Keplerian velocity in the inner disk. {We refer to the 2D map of $v_{\phi}/v_{\rm K}$
in Fig. \ref{fig:2DEvolDisk} for the visual illustration.}
Here, the $\CR$ and $v_{\rm K}$  are presented by the vertical blue dash-dotted line and black dash-dotted line, respectively. 
In our model disk, the super-Keplerian region advances even inward of $\CB$. We notice that when the disk becomes substantially massive and the angular momentum increases, the rotational velocity at the disk's outer edge becomes super-Keplerian across an annulus of the disk where self-gravity of the disk becomes significant. 
This can be further understood from the gravitationally unstable region (Toomre-$Q\lesssim 2$) shown in Fig. \ref{fig:torqueanvel_model2}e, which essentially coincides with the same disk annulus exhibiting super-Keplerian rotation.
Our results show that the region of super-Keplerian rotation begins near $\CR$ and extends further inward of $\CB$.
It can be realized that the model disk remains substantially massive that it induces additional centrifugal acceleration due to the self-gravity effect of the disk  (refer to the middle panel of Fig.~\ref{fig:1Dmodel}). 
This is further confirmed by the mass-weighted azimuthally averaged radial profile of the rotational velocity ($v_{{\phi},{\rm m-wt}}$), which coincides with the same annuli identified as super-Keplerian region in the $v_{\phi}-r$ profile and GI unstable annuli seen in Fig.~\ref{fig:torqueanvel_model2}d and Fig.~\ref{fig:torqueanvel_model2}e, respectively.

Afterwards, Fig.\,\ref{fig:torqueanvel_model2}e shows the radial profile of $Q_{\rm min}$ and $Q_{\rm m-wt}$. The former and latter represent the minimum in the Toomre-$Q$ parameter across all azimuths and mass-weighted azimuthally averaged Toomre-$Q$ , respectively. 
The blue shaded region shows the azimuthal scatter of Toomre-$Q$ at a given radius, spanning the range between its minimum and maximum values. Note that, $Q_{\rm min}$ corresponds to the lower boundary of this shaded band.
Here, the Toomre-$Q$ parameter can be considered as a measure of gravitational instability (GI) in the disk. 
As the protostellar disk grows and becomes massive, it becomes GI unstable.  
The gravitationally unstable region in our model
disk is defined by the Toomre-$Q$ criterion where $Q \lesssim 2$ \citep[][]{Toomre1981,Polyachenko1997}. 
In our analysis, $Q_{\rm min}$ and $Q_{\rm m-wt}$ essentially act as better tracers for the GI in disk as the disk is non-axisymmetric (see  Fig.~\ref{fig:2DEvolDisk}). 
In particular, the $Q_{\rm min}$ traces the densest and most gravitationally unstable azimuth in the disk. 
And the $Q_{\rm m-wt}$ delineates the contribution from the disk's dense regions to the GI activity, and thus less affected by the influence of low-density regions such as the inter-spiral-arm area, which would otherwise increase the local Toomre-$Q$.

\begin{figure}[!htb]
\includegraphics[width=\linewidth]{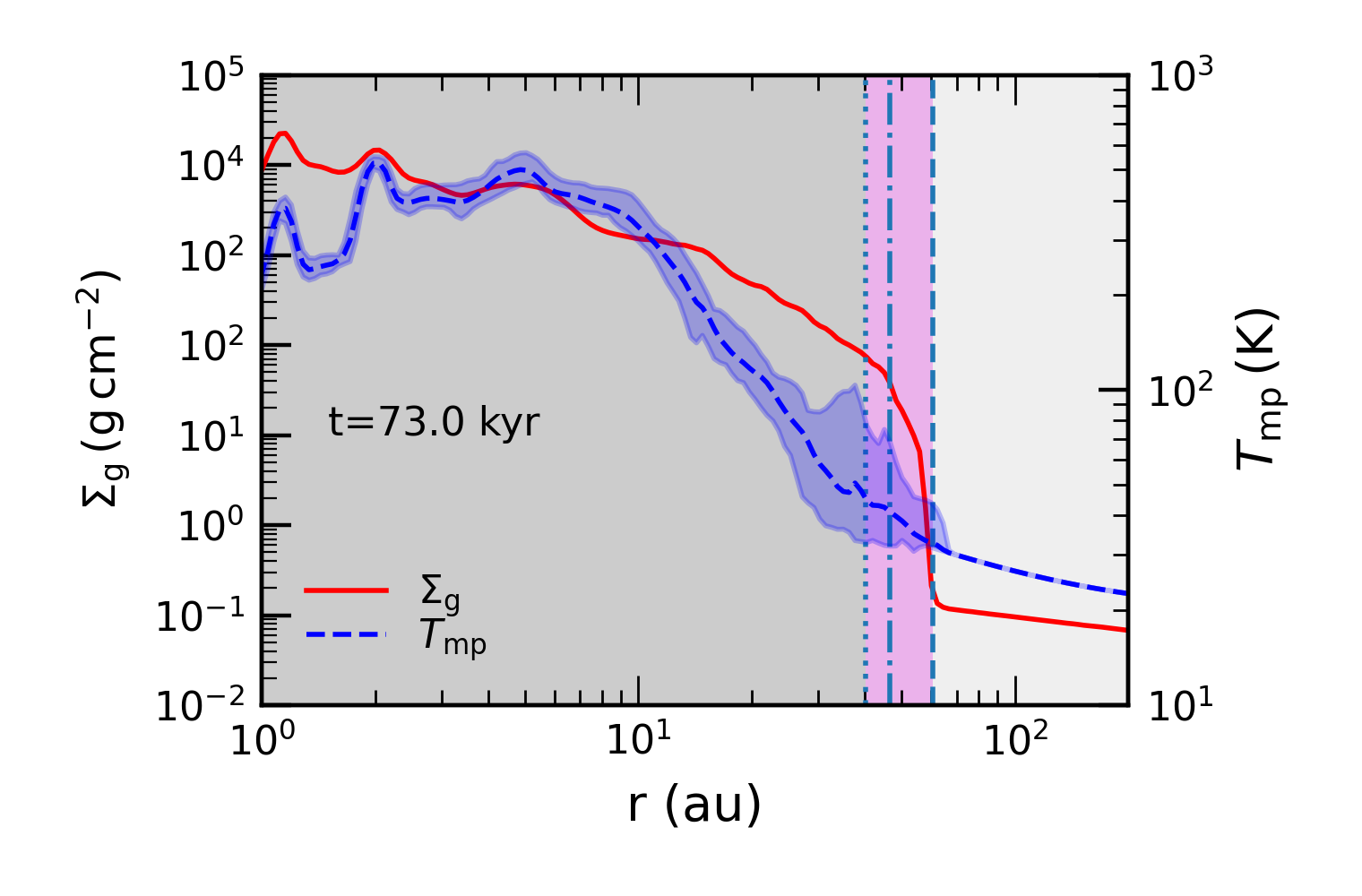}
\vspace*{-0.8cm}
\caption{The azimuthally averaged radial profiles of the gas surface density ($\Sigma_{\rm g}$) and midplane temperature ($T_{\rm mp}$) at a distinct evolutionary time as obtained from \modeltwo. The blue shaded region shows the azimuthal scatter of $T_{\rm mp}$
at a given radius. 
The vertical dashed, dash-dotted, and dotted line represent $\rOTZ$, $\CR$, and $\CB$, respectively. The pink shaded region represents the \EDTZ. The lighter and darker gray shaded regions represent the envelope and the disk interior, respectively. 
}
\vspace*{-0.1cm}
\label{fig:sigmaTemp}
\end{figure}

In Fig.~\ref{fig:torqueanvel_model2}e, we further see that the GI unstable annuli of the disk rapidly grows with time from a narrow annulus around 4-18~au at about 44.2~kyr to a wide region, ranging from  about 4.5-46~au by about 73~kyr. 
It is within these annuli that the disk exhibits super-Keplerian rotation, a manifestation of the GI unstable massive disk in action.
On both inside and outside of this GI unstable annuli, $Q_{\rm min}$ and $Q_{\rm m-wt}$ sharply increase. The increase in $Q_{\rm min}$ and $Q_{\rm m-wt}$ in the outer region, is related to the transition from the disk to rarefied envelope. 
The increase in $Q_{\rm min}$ and $Q_{\rm m-wt}$ in the inner several au is caused by disk warming (refer to Fig. \ref{fig:sigmaTemp}) and high rates of shear.
The mass transport rate by GI is strongest in the intermediate and outer disk regions where the Toomre-$Q$ parameter is lowest and weakens in the inner several au 
\citep[see][for details]{Vorobyov+2024}. 
Our results show that the disk can exhibit a physically significant self-gravity effect if it is substantially massive. 
Mass infall from the envelope onto the disk’s outer regions causes the disk to become massive and drives the disk into gravitational instability. 
The resulting net negative gravitational and viscous torques transport matter inward, balancing the continuing mass influx from the envelope. 
The star-disk system reaches a quasi-equilibrium in which infalling mass is efficiently carried through the disk by gravitational torques. Because the disk remains massive in this state, the rotation curve becomes super-Keplerian.

In \modelone\, and \modelthree, the resulting torques and the kinematical characteristics also qualitatively show the similar trends to those seen in the fiducial model. 
Figure \ref{fig:torqueangvel_model1model3}a,b show the \EDTZ\, characteristics for \modelone\,\, and \modelthree, respectively. 
It essentially presents the respective model quantities, which are azimuthally averaged radial profiles of local gravitational and viscous torques ($\tauG$ and $\tauV$) and the cumulative gravitational and viscous torques ($\CtauG$ and $\CtauV$), specific angular momentum ($j$), rotational velocity ($v_{\phi}$), $Q_{\rm min}$ and mass-weighted Toomre-$Q$ parameter ($Q_{\rm m-wt}$), and infall velocity ($v_r$) at two given time stamps.
Note that, for \modelone~ the behavior of gravitational and viscous torques is similar to that of the fiducial model (refer to Fig.~\ref{fig:torqueanvel_model2}). 
However, for \modelthree,  magnitude of $\tauV$ within the region between $\rOTZ$ and $\CR$ is only about 1/10 of the $\tauG$ at an early evolutionary time $t=44.2\,{\rm kyr}$, however, it becomes negligible within the \EDTZ\, at later times. 
Nevertheless, the development of the 
$j-r$ jump within the \EDTZ\, is predominantly driven by positive gravitational torques for both these models. 
The kinematical profiles of $j$, $v_{\phi}$, $v_{r}$ also follow the similar trends as found in the fiducial model. Finally, the radial profile of $Q_{\rm min}$, $Q_{\rm m,wt}$, and $v_{{\phi},{\rm m-wt}}$ together present the consistent picture of the super-Keplerian region, similar to what is seen in the fiducial model.

Fig. \ref{fig:sigmaTemp} shows the azimuthally averaged radial profiles of the gas surface density ($\Sigma_{\rm g}$) and midplane temperature ($T_{\rm mp}$) at an evolutionary time $t=73$~kyr. 
The blue shaded region shows the azimuthal scatter of $T_{\rm mp}$ at a given radius, spanning the range between its minimum and maximum values. We notice a steep increase in $\Sigma_{\rm g}$ by about three orders of magnitude across the \EDTZ.
Our numerical results show that $T_{\rm mp}$ remains around $20\,{\rm K}$ in the inner envelope and it rises to approximately $40\,{\rm K}$ across the \EDTZ. Further inside the disk, $T_{\rm mp}$ increases from a few tens of Kelvin to up to $\sim$200\,K, primarily due to adiabatic compression and viscous heating in the inner disk regions \citep{Vorobyov+2020a}. 
In the eDisk study, \cite{Hoff+2023} reported that
molecular line observations of $^{13}$CO have shown that the disk around L1527 IRS protostar is warm $(\sim 20-40\,{\rm K})$, with midplane temperatures too high for CO to freeze-out based on the presence of CO gas out to at least around the edge of the disk at $\sim$75~au. 
In our model disk, $T_{\rm mp}$ of gas at the edge of the disk is quite in agreement with the findings of \cite{Hoff+2023}. 
They also reported a potential temperature enhancement at the envelope-disk interface based on their measurements of  $^{13}\mathrm{CO}/\mathrm{C}^{18}\mathrm{O}$ line ratio. 
Although this line ratio may be underestimated to greater missing flux in the more spatially extended $^{13}\mathrm{CO}$ emission relative to that of $\mathrm{C}^{18}\mathrm{O}$, which may require further investigations.
The gradual increase in the gas $T_{\rm mp}$ across our model \EDTZ  may also support the potential temperature enhancement.

Furthermore, Fig. \ref{fig:sigmaTemp} illustrates that within the inner disk of $\sim$10~au, $T_{\rm mp}$ increases above 500\,K-600\,K due to heating of the dead zone owing to residual turbulence and $PdV$ work \citep[see further in][]{Vorobyov+2020a}. 
A study by \cite{Ueda+2022} also reported that 
within the inner dead zone of the disk ($\approx4$~au), $T_{\rm mp}$ can reach $\sim 1000$~K.
It is important to observationally probe whether
gas indeed pile up at the dead zone. 
Other independent studies of the MRI burst mechanism also report the triggering of MRI bursts when the midplane temperature reaches $\sim$1000 K \citep{Zhu2020,Bae2014}.
Accessing the inner disk directly can be observationally challenging due to the requirement of an angular resolution of less than $0.1''$.
Most of the observational constraints from the inner disk
are derived from the near-infrared (NIR) emission, which allows us 
to only probe the very hot (above 1000\,K) and low-density regions
of the uppermost disk atmosphere \citep[e.g.,][]{Lazareff+2017,Gravity+2019}.


\subsection{Observational aspects of the transition zone} \label{sec:obs}
\begin{figure*}
\centering
\includegraphics[width=0.9\linewidth]{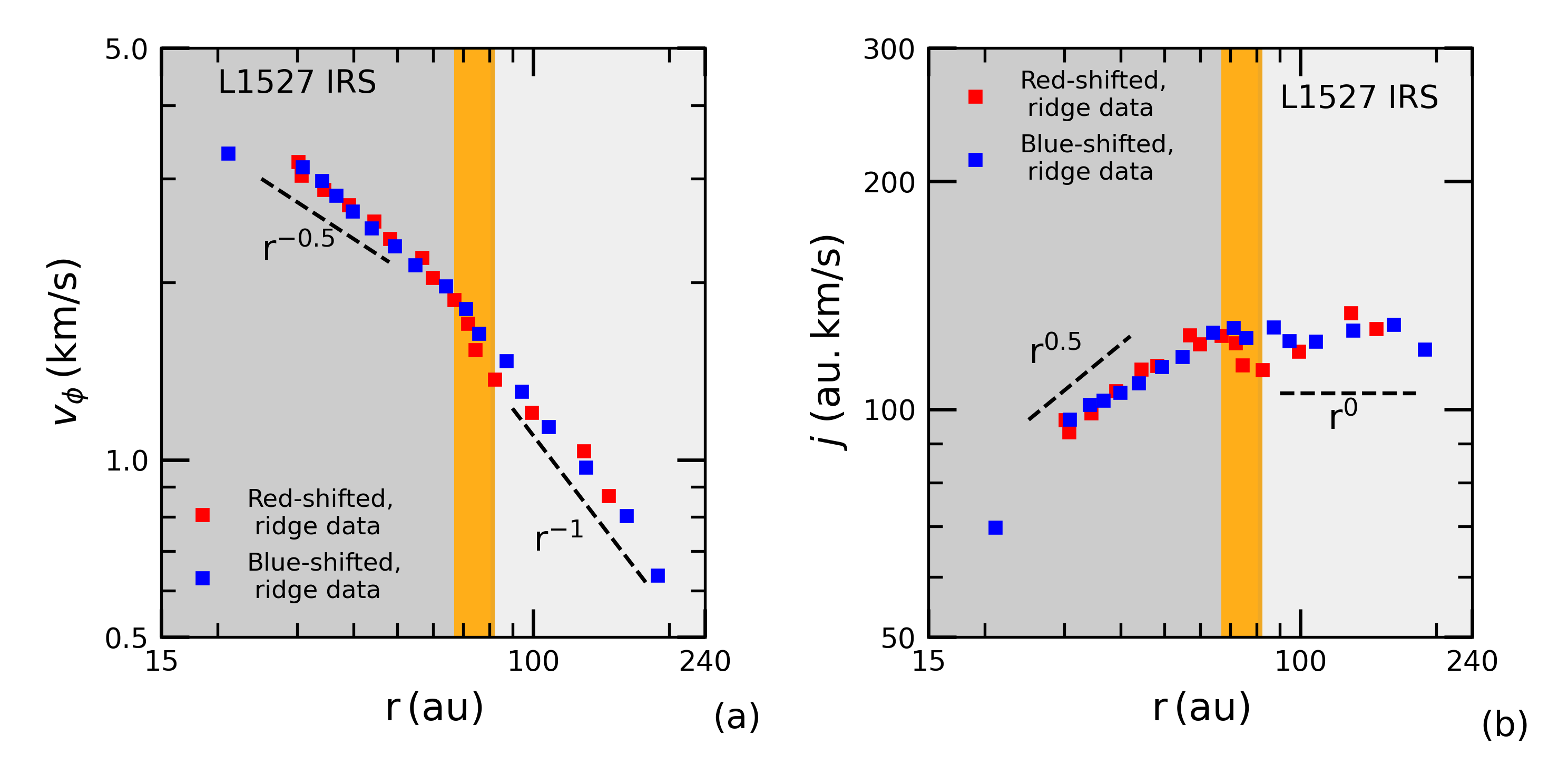}
\vspace*{-0.1cm}
\caption{Radial profiles of 
the rotational velocity ($v_{\phi}$) and specific angular momentum ($j$) in the left (a) and right (b) panel, respectively. 
Both panels show the scattered data points (in blue and red squares) for class 0/I protostar L1527 IRS from the ALMA Large Program  original eDisk data. 
The orange vertical strip represents the jump in $j-r$ profile and its corresponding jump in $v_{\phi}-r$ profile wrt the red-shifted component, as $j$ transitioning from $j\sim {\rm constant}$ in the infalling-rotating envelope (light gray region) to $j\sim r^{1/2}$ in the Keplerian disk (dark gray region)}. 
\label{fig:obseDisk}
\end{figure*}

We re-investigated C$^{18}$O kinematics of class 0/I protostar L1527 IRS at the envelope-disk transition using ALMA Large Program eDisk observations   \citep{Ohashi+2023}. 
Fig. \ref{fig:obseDisk}a presents the observed radial profile of rotational velocity ($v_{\phi}-r$)  of  L1527 IRS. The blue and red squares  represent the redshifted and blueshifted components of the observed $v_{\phi}-r$ profile of L1527 IRS based on the ridge data.  
The $v_{\phi}-r$ data shown in Fig. \ref{fig:obseDisk}a is obtained from the original eDisk ridge data as presented in \cite{Hoff+2023}. We refer to the position-velocity diagram and observational details presented in their study.

In Fig. \ref{fig:obseDisk}b, we present the corresponding $j-r$ diagram calculated from the rotational velocity profile for the first time,  which was not shown in \cite{Hoff+2023}. 
The blue and red squares represent the corresponding redshifted and blueshifted components of the $j-r$ data. 
We identify a jump in this observed $j-r$ profile of L1527 IRS in the redshifted data for the first time as $j$ transitioning from $j\propto {\rm const.}$ ($v_{\phi} \propto r^{-1}$) in the infalling-rotating envelope to $j\propto r^{1/2}$ ($v_{\phi} \propto r^{-1/2}$) in the Keplerian disk. 
We notice that the $j-r$ jump  manifests  slight increase of power-law slope in $v_{\phi}$ from $\sim r^{-1}$ before connecting to $\sim r^{-1/2}$.
It is noteworthy that the jump in the observed $j-r$ profile is more pronounced than in the $v_{\phi}-r$ profile, which allows the jump to be identified in the $j-r$ diagram and then traced back to the corresponding radial zone within the $v_{\phi}-r$ diagram. 
The observed $j-r$ jump in the red-shifted component is marked by the vertical orange strip spanning the radial range $\sim$\,67-82~au where 
$j$ transitions from $j \sim {\rm constant}$ to the radial turnover point where the $j-r$ jump converges into $j\sim r^{1/2}$. 
A corresponding orange strip is overlaid in the $v_{\phi}-r$ plot to identify this radial zone corresponding to the $j-r$ jump. 
We also note that, the $j-r$ jump appear less distinct in the blue-shifted component compared to the red-shifted one, although the $j\propto r^{1/2}$ and $j\propto {\rm const.}$ region are present in the blue-shifted data. 
Moreover, the mismatch between the respective redshifted and blueshifted data points may stem from the absorption due to the (nonaxisymmetric) infall motion in the envelope.

The L1527 IRS hosts an edge-on disk, which allows us to better investigate its rotational motions. 
The eDisk project observed this source at a high angular resolution of 0.1\arcsec, corresponding to $\sim$\,14~au, enabling a detailed study of the disk's rotation.
The C$^{18}$O channel map exhibits a spatial resolution of 2.1 au and a velocity resolution of 0.17\,km s$^{-1}$ as measured from the ridge data traced in the PV diagram \citep[refer to][for further details]{Hoff+2023}. The observed $j-r$ jump in L1527 IRS spanning over a width of $\sim$\,15\,au (refer to  orange vertical strip) is sufficiently resolved in the C$^{18}$O channel map of the original eDisk original data. 
The angular resolution of the C$^{18}$O channel map is  0.17\arcsec x 0.14\arcsec (with a geometrical mean of 0.15\arcsec) and the sensitivity is 1.49 mJy\,beam$^{-1}$. From the PV diagram, it is estimated that the S/N (signal-to-noise) ratio of the C$^{18}$O emission used to measure the rotation curve is at least $\sim$10.
This implies that the relative positional difference that can be measured in the map is at least 0.015\arcsec \,as calculated from the ratio of angular resolution to S/N ratio  \citep[refer to][]{Reid+1988}, which corresponds to a spatial resolution of 2.1 au for the case of L1527 IRS, given the distance is $\sim$\,140\,pc.

In the infalling-rotating envelope region, the best-fit power-law index is close to that expected for angular momentum conservation $v_{\phi}\sim r^{-1}$. 
This suggests that rotation in the envelope is insufficient to support material against the gravity of the central protostar, consistent with its infalling motion \citep[as also reported by][]{Ohashi+2014,Aso+2017,Hoff+2023}. The behavior of the outer $j \sim {\rm const.}$ part of the envelope might depend on the Fourier sampling of the interferometer data, a detailed follow-up study  on L1527 IRS will be pursued in a future work.

From a qualitative comparison with our numerical results, based on the radial behavior of the $j-r$ and $v_{\phi}-r$ profiles, we find that the envelope-to-disk transition proceeds through a jump, also observed in the $j-r$ profile of L1527 IRS. The observed $j-r$ jump indicates the existence of a radially extended transition zone at the envelope–disk interface that may arise from angular momentum redistribution at the disk edge. A tailored numerical modeling of L1527 IRS will be conducted in a future work. 
In the current study, we restrict our discussions to the observed 
$j-r$ jump to remain focused on the main context, which is qualitatively consistent with the radial features found in our simulations. Using the dynamical mass derived from the best-fit Keplerian radius ($\CR$) and assuming the bolometric luminosity equals the accretion luminosity, the inferred mass accretion rate of L1527 IRS is
is about $\sim 3.9\times 10^{-7}\,\Msun\,{\rm yr}^{-1}$ \citep{Tobin+2012,Aso+2026}. Our fiducial model yields a baseline mass accretion of $2-5\times 10^{-6}\,\Msun\,{\rm yr}^{-1}$ , which is qualitatively consistent with that of L1527 IRS (refer to Fig.~\ref{fig:1Dmodel} in Appendix~\ref{sec:AppSimulations}). 
The dynamical mass of the Class 0/I protostar L1527 IRS is estimated at 
$\sim$\,0.3-0.5\,$\Msun$ based on the best-fit $v_{\phi}$ at the $\CR$ w.r.t the C$^{18}$O ridge data \citep{Ohashi+2014,Aso+2017,Hoff+2023,Aso+2026}.  Our model azimuthally averaged 
$j-r$ profile at $t$=76.6 kyr shows the onset of the $j-r$ jump around a similar radial location as of the observed one and this model profile corresponds to a stellar mass of $\sim$ 0.55 $\Msun$ (Fig.~\ref{fig:1Dmodel}), reasonably consistent with the dynamical mass of L1527 IRS. A more quantitative comparison will be pursued in the future study.
Although the model $j-r$ jump magnitude differs from the observed jump, this may result from partial angular momentum loss either via MHD winds and/or magnetic braking, indicating that internal torques can drive the $j-r$ jump, albeit with reduced amplitude. However, consideration of these cases fall beyond the scope of the current study and will be explored in future work.


\section{Discussion}
\label{sec:discussion}
In this study, we investigated the break in the $j-r$ profile at the envelope-disk interface, with an emphasis on the physics of \EDTZ\,  from the MHD disk simulations with self-consistent cloud-to-disk transition (refer to Sec. \ref{sec:theory}, Fig. \ref{fig:torqueanvel_model2}). 
To further strengthen our understanding of the \EDTZ, we  present the observational evidence of the $j-r$ jump in L1527 IRS identified in this study, that can serve as a kinematical signature for the existence of \EDTZ\,  (Fig.~\ref{fig:obseDisk}).
This kinematical tracer of $j-r$ jump in L1527 IRS can offer an independent pathway to probe the detailed kinematics of the transition zone in the future studies.

Our numerical results suggest that the jump in the $j-r$ profile is characterized by a positive local gravitational torque ($\tauG$), implying inward transportation of the angular momentum and outward transport of mass (refer to Fig.~\ref{fig:torqueanvel_model2}). 
As $\tauG$ changes from predominantly positive to negative, when moving radially inward, 
it implies that the predominant outward mass transport within the \EDTZ\,  changes into the predominant inward mass transport in the interior of the disk. 
In this study, the disk formation and  early evolution is primarily regulated by the gravitational torques as they dominate viscous torques in the embedded phase of disk evolution (refer to Fig.~\ref{fig:torqueanvel_model2}a,b; \ref{fig:torqueangvel_model1model3}a,b), which is consistent with the findings of \cite{VorobyovBasu2009}. 
Several comprehensive numerical studies have examined protostellar disk formation and evolution from the magnetized cloud core collapse \citep[e.g.,][etc.,]{Tomida+2015,Zhao+2018,Gray+2018,Wurster+2018} and related kinematical studies of \citep[e.g.,][]{Oya+2015, Oya+2022} also have provided insights on the dynamics within the collapsing envelope and protostellar disk formation. However, our numerical analysis is the first to explicitly present and characterize the  kinematics based on the self-consistent internal torque characteristics 
for determining the physical origin of a radially extended \EDTZ.

Our numerical findings of \EDTZ\, illustrate that the outer edge of the \EDTZ\, ($\rOTZ$) extends beyond $\CR$ by approximately a factor of 1.3-1.7  
(refer to Fig. \ref{fig:torqueanvel_model2}, \ref{fig:torqueangvel_model1model3}). 
We also notice that at $\rOTZ$  infall velocity ($v_r$) begins to sharply decline in magnitude and it drops by roughly 70\%–90\% relative to its value at $\rOTZ$ by the time it reaches at $\CR$. 
In our model disk, the ratio $\CR/\CB \sim$ is estimated to be about 1.1–1.5 (refer to Fig. \ref{fig:torqueanvel_model2}, \ref{fig:torqueangvel_model1model3}).
We note that, the classical ratio $\CR/\CB$ of 2  based on a (pressure less) test particle approach \citep[e.g., as adopted in][]{Sakai+2014} is not supported by our numerical treatment that employs a fluid-based approach including the gas pressure gradient, disk self-gravity, mass and angular momentum transport due to viscosity and disk gravitational instability.

Furthermore, we noticed the super-Keplerian region in our model disk  emerges interior to $\CR$ and extends further  inward of $\CB$, where the disk's self-gravity 
becomes significant and the respective annulus of the disk becomes GI unstable (refer to Fig. \ref{fig:torqueanvel_model2}d,e and discussions in Sec. \ref{sec:theory}). 
In our numerical exploration with a range of $\alpha_{\rm visc}$=$10^{-4}$--$10^{-2}$, we find that global nature of the disk evolution is in line with the conjecture that gravitational torque self-consistently dominate the viscous torque during the embedded phase (refer to Fig.  \ref{fig:torqueanvel_model2}, ~\ref{fig:torqueangvel_model1model3}).
Our study also confirms that the super-Keplerian region arises solely from the disk’s self-gravity and persists throughout the embedded phase for all the models, including for \modelthree~ that corresponds to a  $\alpha_{\rm visc}$ of $10^{-2}$. 
In contrast, \citet{Jones+2022} reported from their study of hydrodynamic simulations that a super-Keplerian region at the outer edge of the disk can appear when the $\alpha$-viscosity is relatively high (of the order of 0.01–0.1) and this super-Keplerian rotation is  absent for a lower viscosity of the order of $\sim 10^{-3}$. This may illustrate that in their models the occurrence of super-Keplerian effect is viscosity-dependent and it appears at the disk's edge only. 
Whereas, in our study it is primarily governed by disk's self-gravity and seen over an extended region of the disk, rather than being a mere disk-edge effect.

The current study is limited to the physics captured by (ideal-) MHD  collapse simulations of a highly supercritical prestellar core, including the effects of self-gravity and gravitational torques. 
Our study shows that the redistribution of specific angular momentum (via gravitational torques) is the key mechanism that makes the jump in the $j-r$ profile appear more defined (or sharpened) rather than smoothening it out. 
Other scenarios may be explored to investigate the physical mechanism of such a transition zone, including the effects of MHD disk winds or magnetic braking, which is beyond the current scope of this study. 
In the presence of MHD disk winds, angular momentum is expected to be lost from the system, rather than being redistributed within the disk, potentially leading to a suppressed or entirely absent jump in the $j-r$ profile across the \EDTZ\, when transitioning from $j\sim {\rm constant}$ to $j\sim r^{1/2}$.
In another scenario of disk evolution, where the angular momentum is lost due to strong magnetic braking, there is a smooth transition (meaning no jump) in the $j-r$ profile when going from the inner envelope to the disk \citep{Tsukamoto2024,Kobayashi+2025}. 
Contemporary numerical studies \citep[e.g.,][]{Krasnopolsky+2011,Li+2011,Tomida+2015,Wurster+2018,Zhao+2020,Lee+2021,Tu+2024} also suggest that magnetic braking seems to be more significant in driving protostellar angular momentum evolution.
Our current scenario may resemble the case where magnetic braking is effectively suppressed due to sufficient Ohmic dissipation, as described in a study by \cite{MachidaBasu2024}. 
The magnetic braking effect is likely significant during the early Class 0 phase and the remaining envelope is available to facilitate angular momentum transport. 
Because of this, it may be possible that the \EDTZ\, in the $j-r$ profile is suppressed or remains absent during the early stages when magnetic braking is still strong \citep{Lopez+2024} or only develops at later times.
More studies are needed to explore the role of large-scale gas kinematics in global long-term nonideal MHD models.

While the $j-r$ 
jump identified in L1527 IRS in our study stands out as a kinematical signature for the presence of an \EDTZ, it is important to understand that radial behavior of $j-r$ may vary among different young stellar objects (YSOs).  
For example, a recent study by \citet{Lin+2025} reported a dip in rotation (and thus a drop in $j$) in H111 at around 600~au, with the Keplerian radius estimated to be 
$\sim$160 au from molecular line observations \citep{Lee+2016}. Additional observational studies of other YSOs focusing on the transition zone will be carried out in future works to investigate the variety of tracers for \EDTZ\, and to interpret their complex characteristics through different physical processes at work, including the internal torques, magnetic braking, and MHD winds, etc.


\section{Conclusions}   \label{sec:conslusions}
We conduct MHD disk simulations with self-consistent cloud-to-disk transition to to investigate the break in the radial profile of specific angular momentum at the envelope-disk transition and to determine the physics of \EDTZ. 
Motivated by our numerical results showing a $j-r$ jump as a tracer of model \EDTZ, we present the observational evidence of a $j-r$ jump feature for the case of class 0/I protostar L1527 IRS within our conceptual framework.
Our key findings are listed as follows:

\begin{enumerate} 
    \item 
    Our numerical results reveal that there exists a jump in $j-r$ and $v_{\phi}-r$ profile, embedded within the \EDTZ, is characterized by positive local gravitational torque, implying angular momentum going inward (refer to Fig. \ref{fig:torqueanvel_model2}a,b). 
    The \EDTZ\, spans between two dynamical regions--- the infalling-rotating envelope 
    following $j \sim {\rm constant}$ ($v_{\phi} \sim r^{-1}$) and the Keplerian disk that follows $j \propto r^{1/2}$ ($v_{\phi} \sim r^{1/2}$),  refer to Fig. \ref{fig:torqueanvel_model2}.
    The redistribution of the angular momentum from inner to outer region of the disk 
    and the positive gravitational torque 
    within the \EDTZ\, adds up to the development of such a jump in the $j-r$ profile at the envelope-disk transition and essentially making it appear more well defined and sharper.

    \item We identify three key radii that are critical for understanding the physics of the model \EDTZ\, (see Fig. \ref{fig:torqueanvel_model2}):
   (i) the outer edge of the transition zone ($\rOTZ$) where 
the infall velocity begins a sharp decline in magnitude and $j$ begins a transition from $j \sim {\rm constant}$ towards $j\sim r^{1/2}$, (ii) 
progressively radially inward, 
the centrifugal radius ($\CR$) where 
$v_{\phi}$ 
first transitions to the Keplerian velocity ($v_{\rm K}$) at the disk's edge, 
(iii) the inner edge of the \EDTZ\, is identified at the centrifugal barrier ($\CB$), where 
the radial infall speed ($v_r$) drops to negligible values.

\item In our model disk, inward
of $\CR$, $v_{\phi}$ transitions to a super-Keplerian value accounting for the disk’s self-gravity, before returning to Keplerian velocity in the inner disk. The super-Keplerian disk region extends even inward of the $\CB$.
In addition, our numerical results (based on the fluid treatment) 
yield a $\CR/\CB$ ratio of 1.1-1.5, in contrast to the classical ratio of 2 based on the classical test particle approach.

\item 
    From the original eDisk observations of L1527 IRS, we identify a jump in the observed $j-r$ profile at the envelope-disk transition that suggests as a kinematical signature for the existence of an \EDTZ\,  (refer to Fig. \ref{fig:obseDisk}) 
\end{enumerate}

Our findings provide a theoretical perspective on the physical origin of a radially extended \EDTZ, supported by the  observational evidence of $j-r$ jump as a kinematical tracer of transition zone, 
 may further offer new insights into angular momentum redistribution in the overall protostellar evolution.

\vspace{-0.3cm}
\begin{acknowledgements} 
We sincerely thank Merel van't Hoff for providing the observational data.
We thank the referee for
the insightful comments.
The authors acknowledge the High-Performance Computation (HPC) support provided by the Digital Research Alliance of Canada (\href{https://alliancecan.ca/en}{alliancecan.ca}) and the regional partner organizations \href{https://www.computeontario.ca/}{Compute Ontario} and \href{https://www.calculquebec.ca/}{Calcul Québec}. The authors also acknowledge the high-performance computing facilities in ASIAA.
I.D. and N.O. acknowledge the support from National
Science and Technology Council (NSTC) in Taiwan through
the grants NSTC 113-2112-M-001-037, and also from Academia Sinica
Investigator Project grant (AS-IV-114-M02). 
S.B. is supported by a Discovery Grant from NSERC. 
E.I.V acknowledges support by the Austrian Science Fund (FWF), project I4311-N27.
\end{acknowledgements}


\appendix
\section{Numerical Methods}
For the convenience of the reader and for completeness,
we summarize the numerical methodology used in our paper. 
\subsection{Numerical Equations for gaseous component} \label{sec:gas}
The equations of mass continuity, momentum conservation, and energy transport are solved in the thin-disk approximation, which can be written as follows
\begin{equation}
\label{eq:cont}
\frac{\partial \Sigma_{\rm g}}{\partial t}  + \vec{\nabla}_{\rm p}  \cdot \left(\Sigma_{\rm g} \vec{v}_{\rm p} \right) = 0,  
\end{equation}
\begin{align}
      \frac{\partial}{\partial t} \left( \Sigma_{\rm g} \vec{v}_{\rm p} \right) + \left[ \vec{\nabla} \cdot \left( \Sigma_{\rm g} \vec{v}_{\rm p} \otimes \vec{v}_{\rm p} \right) \right]_{\rm p}  & =  \notag \\
      - \vec{\nabla}_{\rm p} {\cal P}  + \Sigma_{\rm g} \, \left( \vec{g}_{\rm p} +\vec{g}_\ast \right) 
+ \left(\vec{\nabla} \cdot \mathbf{\Pi} \right)_{\rm p}  & \notag \\
- \Sigma_{\rm d,gr} \vec{f}_{\rm p} +  {B_z {\vec B}_p^+ \over 2 \pi} - H_{\rm g}\, \vec{\nabla}_{\rm p} \left({B_z^2 \over 4 \pi}\right), ~ ~ ~
    \label{eq:mom}
\end{align}

\begin{equation}
\frac{\partial e}{\partial t} +\vec{\nabla}_{\rm p} \cdot \left( e \vec{v}_{\rm p} \right) = -{\cal P}
(\vec{\nabla}_{\rm p} \cdot \vec{v}_{\rm p}) -\Lambda +\Gamma + 
\left(\vec{\nabla} \vec{v}\right)_{pp'}:\Pi_{pp'}, 
\label{eq:energy}
\end{equation}
respectively, where the subscripts $p$ and $p'$
refer to the planar components
$(r, \phi)$ in polar coordinates, 
$\Sigma_{\rm g}$ is the gas surface density, $\vec{v}_{\rm p} = v_r \hat{r} + v_{\phi} \hat{\phi}$ is the gas velocity in the disk plane, $\nabla_{\rm p} = \hat{r} \partial/\partial r + \hat{\phi} r^{-1} \partial/\partial \phi$ is the gradient along the planar coordinates of the disk, 
${\cal P}$ is the vertically integrated gas pressure, 
$e$ is the internal energy per unit surface area, $\Gamma$ and $\Lambda$ represent the cooling and heating rates as discussed later in Eq. (\ref{eq:cooling}) and (\ref{eq:heating}), respectively. 
The ideal gas equation of state is used to calculate the vertically integrated gas pressure, ${\cal P}=(\gamma-1) e$ with $\gamma=7/5$. 
The gravitational acceleration in the disk plane $\vec{g}_{\rm p} = g_r \hat{r}+ g_{\phi} \hat{\phi}$, takes into account self-gravity of disk by solving for the disk gravitational potential
using the Poisson integral \citep[see details in ][]{BT2008, Vorobyov+2024}.
The term $\vec{g}_{\ast}$ is the gravitational acceleration due to the central protostar, which only has a radial component.
The viscous stress $\mathbf{\Pi}$ tensor is defined as 
\begin{equation}
    \mathbf{\Pi}=  2\Sigma_{\rm g} \nu \left({\nabla} \vec{v} -(\nabla \cdot \vec{v})\frac{\vec{1}}{3} \right)
\end{equation}
where $\nu$ is the kinematic viscosity, $\vec{1}$ is the unit tensor, and $\vec{\nabla}v$ is the symmetrized velocity gradient tensor (see \S 2 and Appendix B of \citealt{VorobyovBasu2009} and also in \citealt{VorobyovBasu2010}). 
For the  actual expressions of $\left(\vec{\nabla} \cdot \mathbf{\Pi} \right)_{\rm p} $, $\left[ \vec{\nabla} \cdot \left( \Sigma_{\rm g} \vec{v}_{\rm p} \otimes \vec{v}_{\rm p} \right) \right]_{\rm p}$, and 
$(\vec{\nabla}_{\rm p} \cdot \vec{v}_{\rm p})$ in polar coordinates $(r,\phi)$, we refer to Appendix C of \cite{VorobyovBasu2010}.
Here, $\nu=\alpha_{\rm visc} c_s H_{\rm g}$, where $c_s^2 = \partial \mathcal{P}/\partial H_{\rm g}$ is the effective sound speed of (generally)
nonisothermal gas.
The code is written in the thin-disk limit, complemented by a calculation of the gas vertical scale height ($H_{\rm g}$) using an assumption of local hydrostatic equilibrium in the gravitational field of disk and star, as described in \cite{VorobyovBasu2009}. 
The resulting model has a flared structure (because
the disk vertical scale height increases with radius), which guarantees that both the disk and envelope receive a fraction of
the irradiation energy from the central protostar. 
The term $\vec{f}_{\rm p}$ is the drag force per unit mass between dust and gas, and $\Sigma_{\rm d,gr}$ is the surface density of grown dust \citep[see details in][]{Vorobyov+2018,Vorobyov+2020a}. 
We refrain from discussing the gas–dust interaction physics in detail to remain within the scope of this focused study.

Coming to the magnetic field physics \citep[see also \S 2.3 in][]{Vorobyov+2020a}, the planar components of the magnetic field are ${{\vec B}}_p^+ = {B}_r^+ \hat{r} + {B}_{\phi}^+ \hat{\phi}$ where the `$+$' corresponds to the component at the top surface of the disk. 
Here, $B_z$ is the vertically constant but radially and azimuthally varying $z$-component of the magnetic field within the disk thickness. 
The planar component of the magnetic field at the top surface of the disk is denoted by ${\vec B}_p^+$ and the midplane symmetry is assumed, such that ${{\vec B}}_p^-=- {\vec{B}}_p^+$. 
In Equation (\ref{eq:mom}) the last two terms on the right-hand side are the Lorentz force (including the magnetic tension term) and the vertically integrated magnetic pressure gradient. 
The magnetic tension term arises
formally due to the Maxwell stress tensor, that can be intuitively understood as the interaction of an electric current at the disk surface 
\citep[see also Figure 1 of][]{DasBasu2021}. 
The discontinuity in tangential field component gives rise to surface current while there is no current within the disk. 
In the adopted thin-disk approximation, the magnetic field
and gas velocity have the following nonzero components in
the disk: $\vec{B}= (0, 0, B_z)$ and $\vec{v} = (v_r, v_{\phi}, 0)$, respectively. 
The vertical component of magnetic field is calculated by explicitly solving the induction equation in the ideal MHD regime:  
\begin{eqnarray}
\frac{\partial B_z}{\partial t} = -\frac{1}{r} \left( \frac{\partial}{\partial r}\left(r v_r B_z\right) + \frac{\partial}{\partial \varphi}\left(v_{\varphi}B_z\right)\right),
\label{eq:Bz}
\end{eqnarray}
wherein the advection of $B_z$ is considered. 
The diffusive effects of Ohmic dissipation and ambipolar diffusion are neglected due to high computational cost. The total magnetic field can be written as the gradient of a scalar magnetic potential $\Phi_{\rm M}$ and the planar component of magnetic field ($\vec{B}_p^+$) is calculated by solving the Poisson integrals \citep[see details][]{Vorobyov+2020a} with the source term of $(B_z - B_0)/(2 \pi)$, where $B_0$ is the constant background magnetic field with value of $10^{-5} \, {\rm G}$.

The local gravitational torques $\tau_{\mathrm{grav}}(r,\phi)$ and local viscous torques $\tau_{\mathrm{visc}}(r,\phi)$ are calculated as below
\begin{equation}
\label{eq:tauG}
    \tau_{\mathrm{grav}}(r,\phi) = 
        - m(r,\phi)\, \left.\frac{\partial \Phi}{\partial \phi}\right. \, ,
\end{equation}
and
\begin{equation}
\label{eq:tauV}
    \tau_{\mathrm{visc}}(r,\phi) =  r \, (\nabla \cdot \boldsymbol{\mathit{\Pi}})_{\phi} \, S(r,\phi)\, ,
\end{equation}
where $m(r,\phi)$ is the gas mass in a cell with polar coordinates $(r,\phi)$, $\Phi$ is the gravitational potential, $\boldsymbol{\Pi}$ is viscous
stress tensor,  $S(r,\phi)$ the
surface area occupied by a cell with polar coordinates $(r,\phi)$.

The heating and cooling rates $\Gamma$ and $\Lambda$, respectively are based on the analytical calculations of the radiation transfer in the vertical direction \citep{DongEtal2016}. 
The equation for the cooling rate is
\begin{equation}
    \Lambda = \frac{8 \tau_{\rm P} \sigma_{\rm SB} T_{\rm mp}^4}{1+2 \tau_{\rm P}+1.5 \tau_{\rm R} \tau_{\rm P}} \ ,
    \label{eq:cooling}
\end{equation}
where $T_{\rm mp} = {\cal P} \mu/{\cal R} \Sigma_{\rm g}$ is the midplane temperature, $\mu=2.33$ is the mean molecular weight, $\cal{R}$ is the universal gas constant, and $\sigma_{\rm SB}$ is the Stefan-Boltzmann constant. Here, $\tau_{\rm P}= 0.5 \Sigma_{\rm d, tot} \kappa_{\rm P}$ and $\tau_{\rm R}= 0.5 \Sigma_{\rm d, tot} \kappa_{\rm R}$ represent the Planck and Rosseland optical depths to the disk midplane, where $\kappa_{\rm P}$ and $\kappa_{\rm R}$ are the Planck and Rosseland mean opacities taken from \cite{Semenov+2003} and scaled to the unit mass of dust.  The optical depths in the calculations are proportional to the total dust surface density.

The heating function takes into account the stellar irradiation at the disk surface as well as background blackbody irradiation. The heating function per unit surface area of the disk is 
\begin{equation}
    \Gamma = \frac{8 \tau_{\rm P} \sigma_{\rm SB} T_{\rm irr}^4}{1+2\tau_{\rm P}+1.5\tau_{\rm R} \tau_{\rm P}} \ ,
    \label{eq:heating}
\end{equation}
where $T_{\rm irr}$ is the irradiation temperature at the disk surface and
\begin{equation}
    T_{\rm irr}^4 = T_{\rm bg}^4 + \frac{F_{\rm irr}(r)}{\sigma_{\rm SB}} \ .
\end{equation}
Here, $T_{\rm bg}$ represents the temperature of the background blackbody irradiation, $F_{\rm irr}(r) = L_* \cos \gamma_{\rm irr}/(4\pi r^2)$ is the radiation flux (i.e., energy per unit
surface area per unit time) absorbed by the disk surface at a radial distance $r$ from the central star having a stellar luminosity $L_*$. The $L_*$ is a sum of the accretion luminosity $L_{\rm *,accr} = 0.5 \, GM_* \dot{M}/R_*$ 
arising from the gravitational energy of accreted gas plus the photospheric luminosity $L_{*, {\rm ph}}$ due to gravitational compression and
deuterium burning in the stellar interior. Here, $\dot{M}$, $M_*$, and $R_*$ are the mass accretion rate onto the star, the stellar mass, and the radius of the star, respectively. 
Furthermore, $\gamma_{\rm irr}$ is the incidence angle of radiation that arrives at the disk surface w.r.t. the normal at a radial distance $r$ \citep[see Eq. (10) of \S 2.1 of][]{Vorobyov+2020a}.
The resulting model has a flared structure, wherein the disk vertical scale height increases with radius.
Both the disk and the envelope receive a fraction of the irradiation energy from the central protostar.
The adopted opacities in the numerical model of FEOSAD do not take dust growth into account.



\subsection{Magnetorotational instability, adaptive viscosity, and ionization fraction} \label{sec:app3}
The magnetorotational instability (MRI) acts in the presence of the magnetic field and causes turbulence in the weakly ionized gas in the shearing Keplerian disk \citep{Balbus1998}. 
The effects of the turbulence generated by the MRI can be described by a viscosity \citep{ShakuraSunyaev1973}, and 
turbulent viscosity generated by the MRI is traditionally thought to play a major role for the mass and angular momentum transport in protostellar/protoplanetary disks.

The development of the MRI requires a certain ionization level and the Galactic cosmic rays are the primary source of ionization, which are external to the disk and penetrate the disk from both above and below. The disk in this picture is thought to be accreted through a magnetically layered structure \citep[][]{Gammie1996, Armitage+2001} as shown in Figure 5.2 of \cite{IDthesis2022}, wherein most accretion occurs via the MRI-active surface layers and a magnetically dead zone may be formed at the midplane if the total disk column density exceeds $\sim 200$~g~cm$^{-2}$. The effective $\alpha_{\rm visc}$ averaged over the disk vertical column may reach $10^{-3}-10^{-2}$ in disks with the sustained MRI to match the observed mass accretion rates \citep{1995ApJ...446..741B,1996ApJ...463..656S,Armitage+2001}.

This layered disk model has recently been brought into question, following observations of  efficient dust settling in protoplanetary disks \citep{Rosotti2023} and numerical magnetohydrodynamics simulations with nonideal MHD effects included \citep{BaiStone2013,Gressel2015}. According to these findings, the MRI may be effectively suppressed across most of the disk extent and we set $\alpha_{\rm visc}$ to a low value of $10^{-4}$, meaning that turbulent viscosity provides only residual accretion via a thin surface layer of the disk.

There are, however, situations when the MRI may still play an important role.   If the temperature of the inner disk rises above 1000~K, the alkali metals (e.g., potassium) become thermally ionized, providing sufficient ionization for the MRI activation across the entire vertical column of the disk, which in turn leads to an accretion burst. 
In the numerical setup of FEOSAD, we account for this possibility by considering the adaptive $\alpha$-parameter. The kinematic viscosity $\nu = \alpha_{\rm visc} c_s H_{\rm g}$ is parameterized using the usual \cite{ShakuraSunyaev1973}  prescription. 
To simulate accretion bursts, we redefine $\alpha_{\rm visc}$ as a density weighted average which follows:
\begin{equation}
\label{eq:alphaeff}
    \alpha_{\rm visc} = \frac{\Sigma_{{\rm MRI}} \, \alpha_{{\rm MRI}} + \Sigma_{{\rm dz}} \, \alpha_{{\rm dz}}}{\Sigma_{{\rm MRI}} + \Sigma_{{\rm dz}}}\, ,
\end{equation}
where $\Sigma_{{\rm MRI}}$ is the gas column density of the MRI-active layer and $\Sigma_{{\rm dz}}$ is that of the magnetically dead layer at a given radial distance, so that $\Sigma_{\rm g} = \Sigma_{{\rm MRI}} +\Sigma_{{\rm dz}}$. 
The values of $\alpha_{\rm MRI}$ and $\alpha_{{\rm dz}}$ in Equation~(\ref{eq:alphaeff}) correspond to the strength of the turbulence in the MRI-active layer and the MRI-dead  layer, respectively. 
Across the disk, the constant background turbulent viscosity $\alpha_{\rm MRI}$ is set equal to $10^{-4}$, $10^{-3}$, $10^{-2}$ for \modelone, \modeltwo, and \modelthree, respectively, as mentioned in Table~\ref{tab:table}. However, if the thermal ionization exceeds $x_{\rm th}=10^{-10}$, which occurs during the MRI burst, $\alpha_{\rm MRI}$ is then set to 0.1 (labeled it as $\alpha_{\rm MRI}^{\rm Burst}$). 
The latter value is motivated by three-dimensional numerical hydrodynamics simulations of the MRI triggering \citep{Zhu2020}. 
Because of the nonzero residual viscosity arising from hydrodynamic turbulence driven by the Maxwell stress in the active layer, a very small value of $10^{-5}$ is considered for $\alpha_{{\rm dz}}$ \citep[][]{OkuzumiHirose2011}. 
The depth of the dead zone in terms of the $\alpha_{\rm visc}$-parameter can be now
determined by the balance between $\Sigma_{\rm MRI}$ and $\Sigma_{\rm dz}$ using Eq. (\ref{eq:alphaeff}).
The value of $\Sigma_{{\rm MRI}}$ is not set to be constant as in many studies of the MRI bursts \citep{Bae2014,Kadam+2022}, but is defined as
\begin{eqnarray}
\label{eq:zeroDZ}
          \Sigma_{\rm MRI} &=& \Sigma_{\rm g}, \, \, {\rm if} \ \Sigma_{\rm g} < \Sigma_{\rm crit} \, , \\ 
     \Sigma_{\rm MRI} &=& \Sigma_{\rm crit},  \, \,    
    {\rm if} \ \Sigma_{\rm g} \geq \Sigma_{\rm crit} \, , 
\end{eqnarray}
where the critical gas surface density $\Sigma_{{\rm crit}}$  for the MRI development is obtained by equating the timescale of MRI growth to the timescale of MRI damping due to Ohmic dissipation \citep[see details in][]{Vorobyov+2020a}, and is expressed as 
\begin{equation}
    \Sigma_{{\rm crit}} = \left[\left(\frac{\pi}{2}\right)^{1/4}\frac{c^2m_{\rm e}\langle\sigma v\rangle_{\rm en}}{e^2}\right]^{-2}B_z^2 H_{\rm g}^3 x^2 \ ,
    \label{eq:DZ}
\end{equation}
where $e$ is the charge of the electron, $m_{\rm e}$ is mass of the electron, $\langle\sigma v\rangle_{\rm{en}}=2\times 10^{-9}\,\mbox{cm}^3\,\mbox{s}^{-1}$ is the slowing-down coefficient \citep{spitzer1978} for the electron-neutral collisions, and $c$ is the speed of light. 

The ionization fraction $x=n_e/n_n$ is determined from the balance of collisional, radiative recombination, and recombination on dust grains. This is expressed as
\begin{equation}
(1-x)\xi = \alpha_{\rm{r}} x^2n_{\rm{n}} + \alpha_{\rm{d}} xn_{\rm{n}} \, ,
\label{eq:ion}
\end{equation}
where $\xi$ is the ionization rate that is composed of a cosmic-ray ionization rate and the ionization rate by radionuclides \citep{UmebayashiNakano2009}, $n_{\rm n}$ is the number density of neutrals, $\alpha_{\rm{d}}$ is the total rate of recombination onto the dust grains, and $\alpha_{\rm{r}}$ is the radiative recombination rate having a form $\alpha_r = 2.07 \times 10^{-11} \, T^{-1/2} \, {\rm cm}^3 {\rm s} ^{-1}$ \citep{spitzer1978}. 
In the regions, where the gas temperature exceeds several hundred Kelvin, an additional term is added to the ionization fraction $x$ from Eq. \ref{eq:ion}, which is the thermal ionization calculated by considering the ionization of potassium, the metal with the lowest ionization potential. 
The cosmic abundance of potassium set to $10^{-7}$ for these calculations. For further details, including the calculation of the total recombination rate on dust grains $a_{\rm d}$, we refer to  \citet{Vorobyov+2020a}. 
Equation~(\ref{eq:zeroDZ}) demonstrates that a sharp increase in $\Sigma_{\rm crit}$ triggers the burst if the ionization
fraction $x$ experiences a sharp rise as well.


\subsection{Initial Conditions} \label{sec:app1}

The initial profile of the gas surface density $\Sigma_g$ and angular
velocity $\Omega$ of the prestellar core has the following form: 
\begin{equation}
\Sigma_{\rm g} = \frac{r_0 \Sigma_0}{\sqrt{r^2+r_0^2}} \ ,
\end{equation}
\begin{equation}
    \Omega = 2\Omega_0 \left(\frac{r_0}{r}\right)^2 \left[\sqrt{1+\left(\frac{r}{r_0}\right)^2} -1 \right] \ ,
\end{equation} 
consistent with an axially symmetric core collapse  \citep{Basu1997}. 
Here, $\Sigma_0$ and $\Omega_0$
are the gas surface density and angular speed at the center of the core, $r_0 = \sqrt{A} c_s^2/ (\pi G \Sigma_0)$ is the radius of the central plateau, where $c_s$ is the local sound speed in the core, $r$ is the radial distance from the center. 
The dimensionless parameter $A$ corresponds to the density perturbation and it is set to $2$ that makes the core unstable to collapse \citep[see][for more details]{Vorobyov+2020a}. 
Cloud cores are also characterized by the ratio of
rotational to gravitational energy $\beta=E_{\rm rot}/|E_{\rm grav}|$, where the
rotational and gravitational energies are calculated as
\begin{equation}
E_{\rm rot}= 2 \pi \int \limits_{r_{\rm in}}^{r_{\rm
out}} r a_{\rm c} \Sigma_{\rm g} \, r \, dr, \,\,\,\,\,\
E_{\rm grav}= - 2\pi \int \limits_{r_{\rm in}}^{\rm r_{\rm out}} r
g_r \Sigma_{\rm g} \, r \, dr.
\label{rotgraven}
\end{equation}
Here, $a_{\rm c} = \Omega^2 r$ is the centrifugal acceleration; $r_{\rm out}$ and $r_{\rm in}$ are the outer and inner cloud core radius. 
The adopted values of $\beta$ lie within
the limits applicable for dense molecular cloud cores, $\beta=(10^{-4} - 7\times10^{-2})$ \citep[as inferred by][]{Caselli+2002}. 
We refer to Table \ref{tab:table} for the initial values for the model parameters.
The gas temperature of the initial prestellar core is set to $T_{\rm}= 20$ K and a uniform background vertical magnetic field strength is taken to be $B_0 = 10^{-5} \, {\rm G}$. 
The initial prestellar cores with a supercritical dimensionless mass-to-flux ratio, i.e., $\mu >1$, can form through ambipolar diffusion (neutral-ion drift) driven gravitational collapse. 
The spatially uniform dimensionless mass-to-magnetic flux ratio $\mu = (2 \pi \sqrt{G})\Sigma_{\rm g}/B_z$ is set equal to 10.0 in this study, and it stays constant during the entire disk evolution in the adopted ideal MHD limit. 


\subsection{Boundary Conditions}\label{sec:app2}
\begin{figure}
    \centering
    \includegraphics[width=0.8\linewidth]{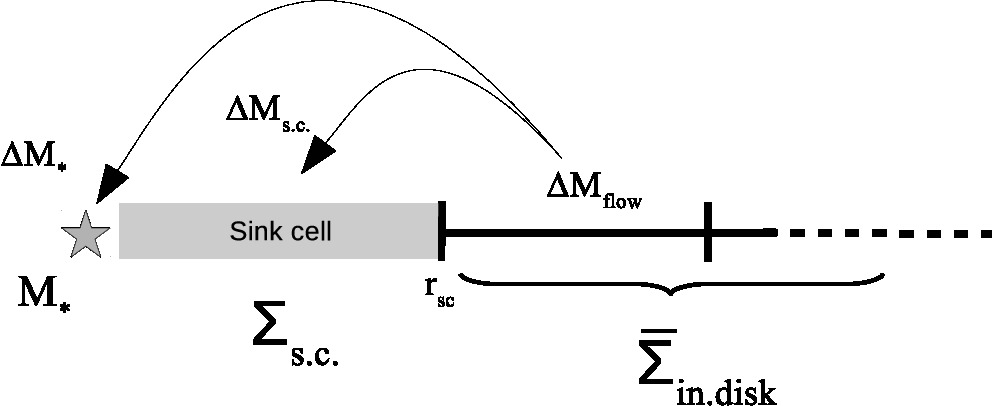}
    \caption{Schematic illustration of the inner boundary condition. The
mass of material $\Delta M_{\rm flow}$ that passes through the sink cell from the active
inner disk is further divided into two parts: the mass $\Delta M_*$  contributing
to the growing central star, and the mass $\Delta M_{\rm s.c.}$ settling in the sink cell, $\Sigma_{\rm s.c}$ is the surface density of gas/dust in the sink cell, $\bar{\Sigma}_{\rm in.disk}$ the averaged surface density of gas/dust in the inner active disk (the averaging is usually done over several au immediately
adjacent to the sink cell, the exact value is determined by numerical experiments); figure is adapted from \citep{Vorobyov+2018}.}
\label{fig:BCs}
\end{figure}

The inner boundary in FEOSAD represents a circular sink cell with a radius of 0.52~au, refer to Fig. \ref{fig:BCs}.  Placing the inner boundary much farther out (at several au) could eliminate the part of inner disk that may be dynamically important since it is where the GI-induced MRI outbursts take place. 
The mass exiting the inner disk is divided between the sink cell, the star, and the jet, with the ratio set to $5\%:85.5\%:9.5\%$. It means that most of the matter quickly lands on the star after crossing the sink-disk interface, but smaller amounts are ejected with the jets and kept in the sink cell. In this realization, the sink cell is characterized by its own surface density of gas and dust, which allows setting an inflow-outflow boundary condition, wherein the matter is allowed to flow freely from the disk to the sink cell and vice versa.  This type of boundary condition allows reducing an artificial drop in the gas density at the inner disk edge, which often forms in outflow-only boundary conditions because of the lack of compensating back flows during wave-like motions triggered, for example, by the spiral density waves or Coriolis force. The angular velocity at the inner boundary is extrapolated according to the Keplerian pattern of rotation and the radial velocity is set equal to the corresponding value of the innermost active disk cell. The flow of matter to and from the sink cell also carries magnetic flux, hence, the inner boundary condition also modifies the vertical component of magnetic field $B_z$ based on the amount of magnetic flux transported. The outer boundary of the computational domain is taken as standard free outflow, where the material is only allowed to leave (no inflow) the computational domain. More details can be found in \citet{Vorobyov+2018} and \citet{Vorobyov+2020b}.


\section{Supplementary Numerical Analysis}
\label{sec:AppSimulations}

\begin{figure}[!htb]
\includegraphics[width=\linewidth]
{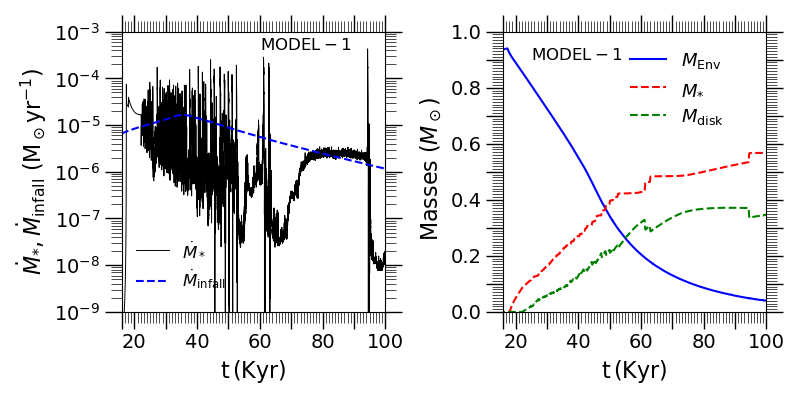}
\includegraphics[width=\linewidth]
{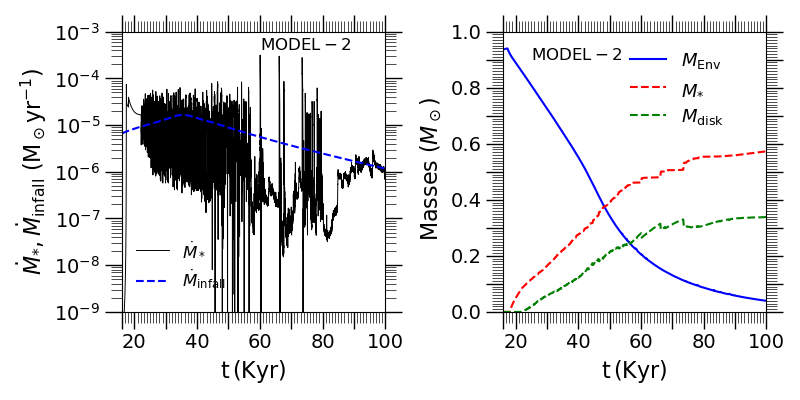}
\includegraphics[width=\linewidth]
{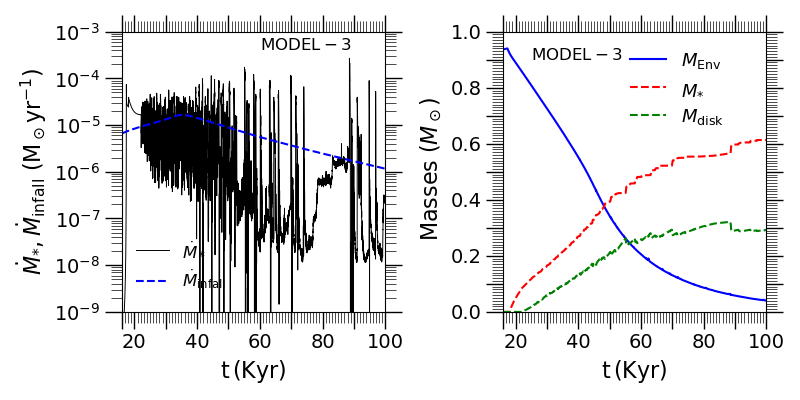}
\vspace*{-0.5cm}
\caption{Temporal evolution of mass accretion quantities at the sink-disk interface (from top to bottom) for \modelone, \modeltwo, and \modelthree\,, respectively. 
Left panel: Mass accretion rate onto the star ($\dot{M}_{\ast}$) and mass infall rate from envelope ($\dot{M}_{\rm infall}$) onto the disk. Right panel: Time evolution of the envelope mass ($M_{\rm env}$), star mass ($M_{\ast}$), and disk mass ($M_{\rm d}$).
}
\label{fig:1Dmodel}
\end{figure}

Figure \ref{fig:1Dmodel} presents the temporal evolution of the mass accretion rate and the mass evolution of the envelope, disk and star for \modelone, \modeltwo, and \modelthree~ in the left and middle panel. 
The mass accretion history is characterized by frequent MRI accretion bursts with magnitude $\ge 10^{-4}~M_\odot$~yr$^{-1}$ superimposed on the quiescent rate of $\approx 10^{-8}-10^{-7}~M_\odot$~yr$^{-1}$.
The accretion bursts are more frequent in the early stages of evolution for all the three models. 
The mass accretion rate is highly variable over time, and in particular, in the early evolution because FEOSAD inherently includes MRI accretion bursts.
In FEOSAD's numerical setup, when the heating of the dead zone owing to residual turbulence and $PdV$ work can raise the gas temperature above 1000~K \citep{Gammie1996,Desch+Turner2015},
thermal ionization of the alkaline metals enables the vigorous MRI growth across the dead zone, followed by a rapid transport of the inner disk material during an episodic GI-driven MRI-triggered accretion burst, which is associated with a turbulent viscosity of $\alpha_{\rm visc}$ = 0.1 in our numerical model \citep[][and see further in Appendix \ref{sec:app3}]{Armitage+2001,Zhu+2009,Vorobyov+2020b,Kadam+2022}.
In \modelone, the bursts cease relatively early, with a prolonged quiescent period interrupted by occasional bursts at later times.
In \modeltwo, the burst activity persists longer and ceases only at about 80~kyr, later stage compared to \modelone~ case. 
In contrast, in the \modelthree~\,, the burst activity continue throughout the timeline of 100~kyr without showing a prolonged quiescent interval as compared to other two models, owing to higher viscous accretion adopted in this model. 
Finally, the temporal evolution of the masses is broadly similar, notably with the $M_{\rm d}$ and $M_{\ast}$ in \modelthree~ growing roughly 14\% less and 7\% more as compared to the other two models by 100~kyr, due to the higher viscous accretion. 
The envelope loses about 50\% and 80\% of its initial mass, at time $t$=44.2 kyr and 73 kyr, respectively.

\begin{figure}[!htb]
\centering
\includegraphics[width=\linewidth]{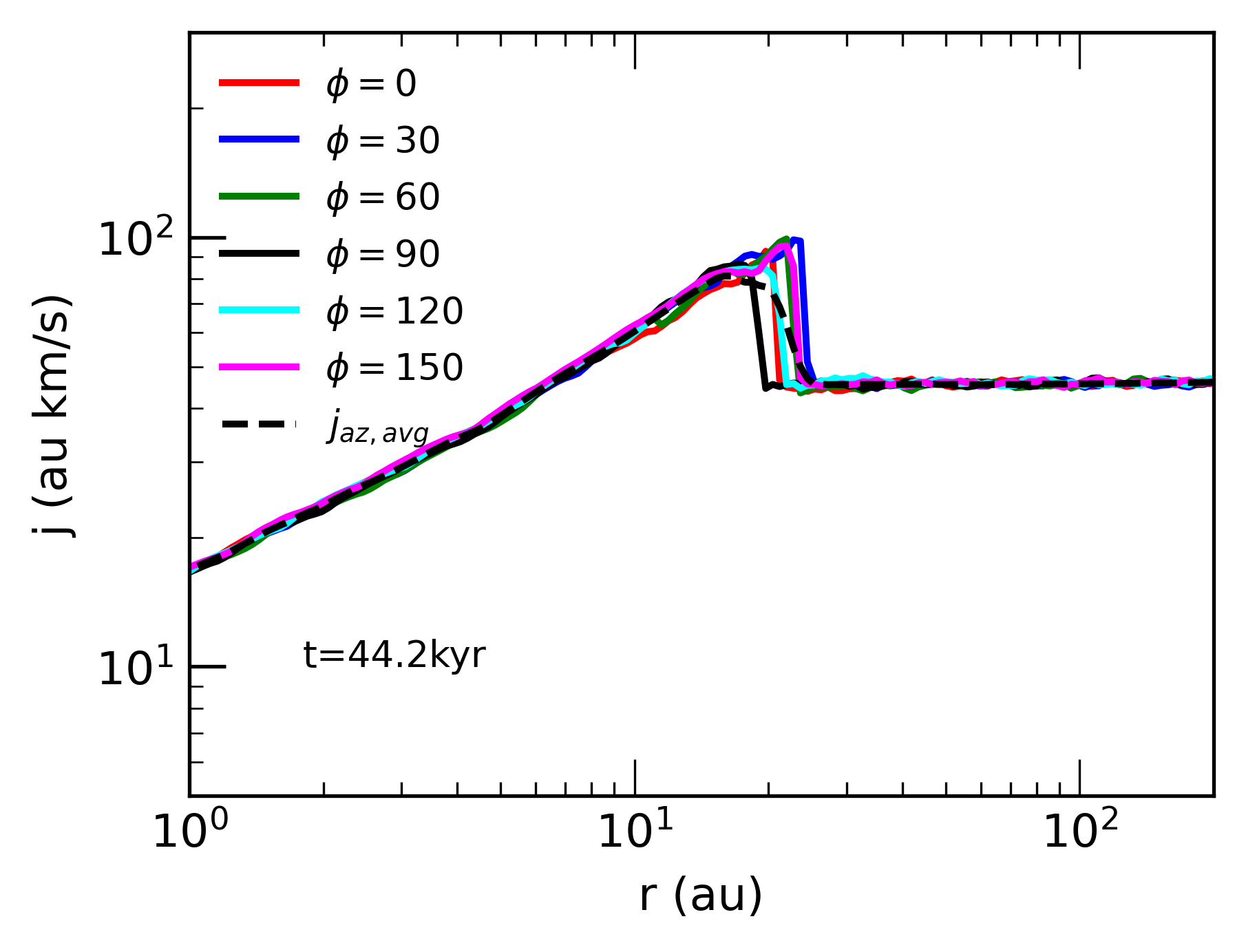}
\includegraphics[width=\linewidth]{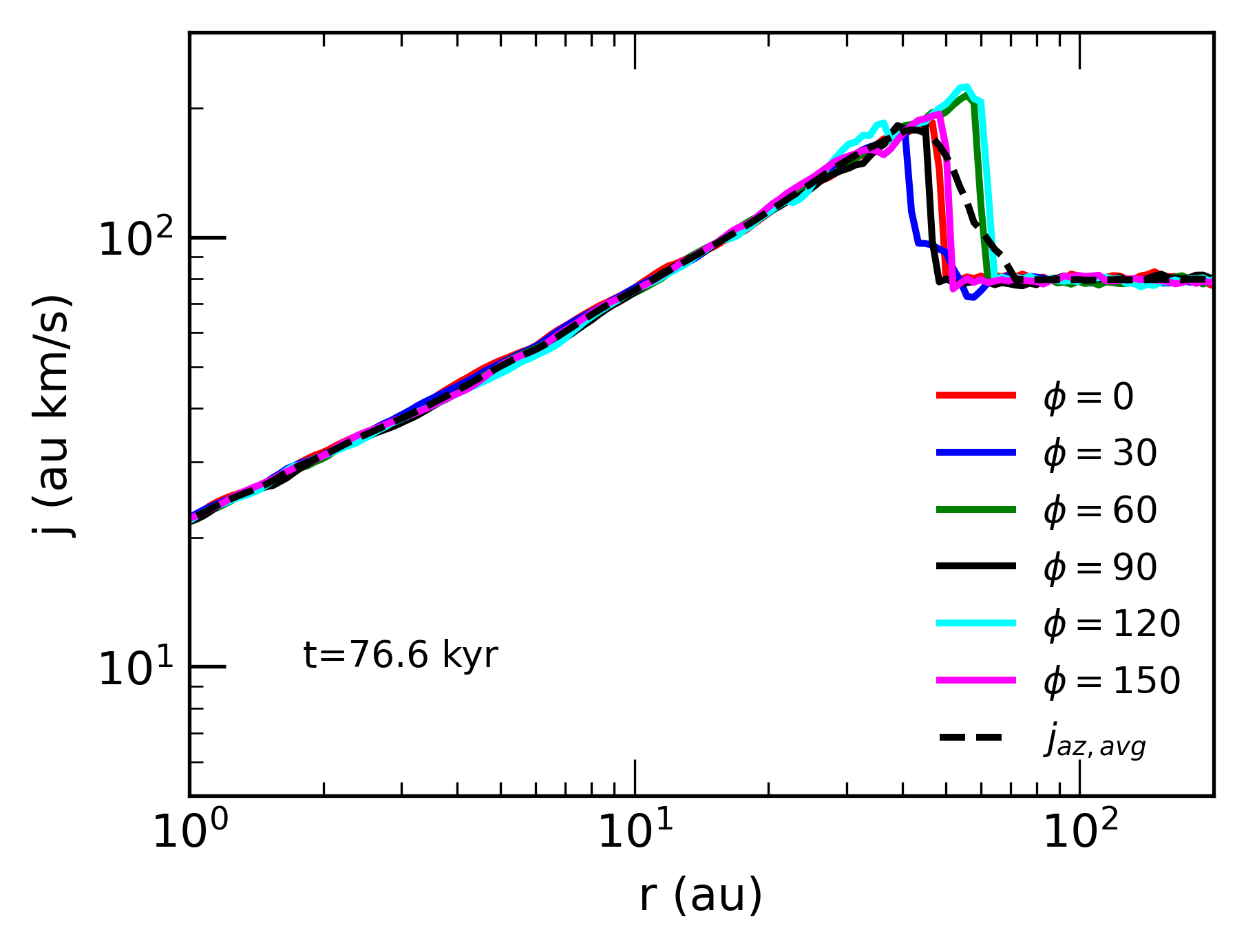}
\vspace*{-0.2cm}
\caption{Radial profiles of the specific angular momentum at different polar rays at two distinct evolutionary times as obtained from \modeltwo. The black dashed curve presents the azimuthally averaged $j-r$ profile as a reference.}
\vspace*{-0.0cm}
\label{fig:jrpolray}
\end{figure}

Here, Fig. \ref{fig:jrpolray} shows the radial profiles of $j$ at different polar rays along with the reference line of azimuthally averaged profile shown by black dashed curve. It shows that $j-r$ jump exists at different polar rays and, in fact, is much sharper.
The azimuthal averaging makes the appearance of the $j-r$ jump a bit smoother. 
At $t$=76.6~Kyr, the azimuthally averaged $j-r$
profile obtained from \modeltwo~ shows the onset of the $j-r$ jump qualitatively at a similar radial location as observed in L1527 IRS.

\bibliography{main}{}
\bibliographystyle{aasjournal}

\end{document}